\documentclass{jfm}
\usepackage[utf8]{inputenc}
\usepackage{natbib}
\usepackage{lmodern}
\usepackage{pifont}
\usepackage{bm}
\usepackage{frcursive}
\usepackage{amsmath,etoolbox} 
\usepackage{dsfont}
\usepackage{url}
\usepackage{textcomp}
\usepackage{soul}
\setstcolor{red}
\usepackage{microtype}
\usepackage{url}
\usepackage{setspace} 
\usepackage{boldline}
\usepackage{ragged2e}
\usepackage{mathtools}

\newtheorem{remark}{Remark}

\definecolor{link}{rgb}{0.18,0.25,0.78}
\usepackage{hyperref}
 \hypersetup{
   colorlinks = true,
  citecolor = link,
   linkcolor = link,
   urlcolor = blue}
\allowdisplaybreaks[1]

\newcommand{\dd}[1]{\:\mathrm{d} #1}

\newcommand{\Lin}[0]{\mathcal{L}}

\newcommand{\ci}[0]{\text{i}}
\newcommand{\ee}[1]{\text{e}^{#1}}
\renewcommand{\Re}{\text{Re}}
\renewcommand{\Im}{\text{Im}}

\definecolor{fgreen}{rgb}{0.13, 0.54, 0.13}

\newcommand{\rev}[1]{\textcolor{black}{#1}}

\title{Internal tide generation from non-uniform barotropic body forcing}
\author{Christos E.\ Papoutsellis\aff{1}, Matthieu J.\ Mercier\aff{1}, Nicolas Grisouard\aff{2} \corresp{\email{nicolas.grisouard@utoronto.ca}}}

\affiliation{\aff{1}Institut de M\'ecanique des Fluides de Toulouse (IMFT), Universit\'e de Toulouse, CNRS, Toulouse,
France
\aff{2}University of Toronto, Department of Physics, Toronto, Canada}

\begin{document}
\maketitle
\begin{abstract}
We model linear, inviscid, internal tides generated by the interaction of a barotropic tide with one-dimensional topography. Starting from the body-forcing formulation of the hydrodynamic problem, we derive a Coupled-Mode System (CMS) using a local eigenfunction expansion of the stream function. For infinitesimal topography, we solve this CMS analytically, recovering  the classical Weak Topography Approximation (WTA) formula for the barotropic-to-baroclinic energy conversion rate. For arbitrary topographies, we solve this CMS numerically. The CMS enjoys faster convergence with respect to existing modal solutions and can be applied in the subcritical and supercritical regimes for both ridges and shelf profiles. We show that the non-uniform barotropic tide affects the baroclinic field locally over topographies with large slopes and we study the dependence of the radiated energy conversion rate on the criticality. We show that non-radiating or weakly radiating topographies are common in the subcritical regime. We also assess the region of validity of the WTA approximation for the commonly used Gaussian ridge and a compactly supported bump ridge studied here for the first time. Finally, we provide numerical evidence showing that in the strongly supercritical regime the energy conversion rate for a ridge (resp. shelf) approaches the value obtained by the knife-edge (resp. step) topography. 
\end{abstract}

\section{Introduction}
Internal tides (ITs) are internal waves generated in the interior of a stratified and rotating ocean through the interaction of the astronomically induced barotropic tidal flow with the seafloor. They are oscillatory perturbations of the baroclinic flow at the tidal frequency, propagating away from bottom irregularities such as ridges or continental shelves \citep{Garrett_Kunze_2007}. ITs are considered to be one of the main sinks of energy for the barotropic tide, and an important contributor to ocean mixing on a global scale \citep{Garrett2003, Wunsch_Ferrari, Whalen2020}. An accurate description of the IT flow and the associated barotropic-to-baroclinic energy conversion is necessary for reliable ocean and climate modelling. Climate models often implement internal-wave-driven mixing through parameterizations involving idealised estimations of the local energy conversion rate at the IT generation sites. Moreover, the dissipation of ITs is also parameterized by using a modal description of the flow \citep{Klymak,CMP2017}.

The problem of IT generation, even in its linearised time-harmonic version considered herein, is quite complex. Simplifications are commonly based on smallness assumptions on the relative topographic height or the so-called criticality defined as the ratio of maximum topographic slope and the characteristic slope of internal waves at the tidal frequency \citep{Garrett_Kunze_2007}. 

\cite{Bell75} linearised the bottom boundary condition around a flat bottom and introduced a uniform barotropic background flow, that is, a barotropic flow that does not depend on the variable topography. This approach, also known as the ``Weak Topography Approximation" (WTA) is formally valid for topographic features of small relative height and criticality \citep{Llewellyn2002, Khatiwala2003, Vlasenko}. Another idealised configuration of opposing nature is to consider discontinuous (infinitely steep) topographies such as one with zero elevation everywhere except at a single point (``knife edge") \citep{LARSEN1969411,llewellyn_smith_young_2003, StLaurent2003, Nycander2006}, \rev{or top-hat, linear ramp  and step functions  \citep{PRINSENBERG, NEW1988691,StLaurent2003}}. An attractive feature of the above simplifications is that they \rev{can lead to} computationally inexpensive \rev{methods of calculating}  the radiating energy, which can be applied on a global scale\rev{, see e.g., } \cite{Nycander2005, Falahat2014}.

The first solution to the full 2D linear problem is due to \cite{Baines1973}, who proposed what is now called the body-forcing formulation. In this approach, the baroclinic flow appears as a response to a non-uniform barotropic flow. That is, a flow with horizontal and vertical velocities that depend on the variable topography and its slope. The spatial part of the response stream function is governed by a hyperbolic partial differential equation (PDE) with variable forcing and homogeneous boundary conditions at the natural boundaries. \cite{Baines1973} proposed a numerical technique based on an integral equation derived from the normal form of this PDE and calculated the radiated energy for simple topographies. He noted, however, that this technique becomes rather involved for complex topographies and proposed in \cite{BAINES1982} a perturbative method for small-scale topographies. \cite{Gerkema2004} obtained numerical solutions of \cite{Baines1973}'s formulation in the time domain by treating the radiation conditions with sponge layers.  \cite{Garett_Gerkemma_2007} showed that the body-forcing term in \cite{Baines1973}'s formulation is inconsistent with non-hydrostatic conditions and derived a consistent formulation. We also adopt this formulation. 

An equivalent formulation is also possible where the governing hyperbolic PDE on the total flow is homogeneous and the forcing appears as a non-homogeneous Dirichlet condition on the bottom boundary (boundary forcing approach) \citep{Sandstrom, Stashchuk, Vlasenko}; see also \cite{Garett_Gerkemma_2007} for the relation with the body-forcing formulation. The numerical techniques developed for this formulation are also based on the normal form of the PDE and its characteristics. They are rather technical, and the treatment of supercritical topographies requires additional attention. Moreover, in order to obtain the flow fields in the physical domain, an additional transformation is required.

Semi-analytical methods have also been developed by using the Green's function of free internal waves corresponding to a radiating source in a flat-strip domain \citep{Robinson1969, Petrelis_et_al_2006, echeverri_peacock_2010, Mathur2016}. By construction, this approach excludes shelf and trench geometries and has been applied only to ridges. Moreover, it is characterised by numerical singularities associated with the solution of an integral equation, as we develop in Section~\ref{sec:discussion}. \cite{Balmforth_Neil_Peacock_2009} developed a variant of this approach in the infinite-depth case with lateral periodic conditions. 

In this work, we develop a new semi-analytical IT model based on \cite{Garett_Gerkemma_2007}'s body-forcing formulation. The analytical step is the exact reformulation of the hydrodynamic problem as an infinite Coupled-Mode System (CMS) of equations accomplished by means of an exact, local eigenfunction expansion of the stream function. This approach, also called coupled-mode theory, has been applied to various non-uniform waveguide problems in acoustics \citep{Brekhovskikh, Desaubies, Maurel,Ivansson}, elasticity \citep{MAUPIN1988, Pagneux, 8805310} and water waves \citep{porter_staziker_1995, athanassoulis_belibassakis_1999, PAPOUTSELLIS2018199} among other disciplines. In the context of ITs, \cite{griffiths2007} derived a CMS from Euler's equations using a local vertical mode decomposition  and calculated 2D ITs over a shelf topography. Similar systems are derived in \cite{Kelly2016} and \cite{lahaye2020}. The principal difference with the present approach is that we work with the stream function. This has the advantage that every term in our local modal expansion satisfies exactly the bottom boundary condition for arbitrary topography, and the solution enjoys faster convergence. We use this CMS to calculate the flow and the barotropic-to-baroclinic energy conversion rates for two types of ridges for a wide and finely resolved range of maximum slopes and heights. To our knowledge, the present work is the first attempt to perform such an extensive set of calculations in the context of the body-forcing formulation.   

The paper is organised as follows. In Section \ref{sec:governing}, we present the body-forcing formulation and derive its energy balance equation. In Section \ref{sec:CMS}, we introduce the modal decomposition of the stream function and derive our CMS. In Section \ref{sec:Validation}, we present numerical convergence and accuracy results, and in Section \ref{sec:fields} we visualise our solutions. In Section \ref{sec:energy}, we consider the conversion rates for various topographies. In Section \ref{sec:discussion} we discuss the differences of the proposed CMS with the Green's function method and, finally, in Section \ref{sec:conclusions}, we present our conclusions.

\section{Governing Equations\label{sec:governing}}

\subsection{Posing the problem\label{subsec:posing}}
In a two-dimensional Cartesian coordinate system $Oxz$, with the vertical axis $z$ pointing upward, we consider a horizontally infinite layer of a density-stratified fluid bounded from above by the ocean surface, modelled as a ``rigid lid'' $\{z=0\}$, and from below by the impermeable bottom $\{z=-h(x)\}$ with $h>0$. We assume that the topography is asymptotically flat, that is, its slope vanishes at infinity, $\lim_{x\to\pm\infty}[h_x] =0$, and we define $\lim_{x\to\pm\infty}h = h_\pm$. The latter requirement allows us to take into account fluid domains with different depths at infinity. The parameters associated with the topography are the characteristic depth $h_0$, the characteristic height $\Lambda$ and the characteristic horizontal scaling length $L$. We do not limit ourselves to small heights; i.e., $\Lambda$ does not have to be much smaller than $h_0$ or any other characteristic vertical length scale. 

In static equilibrium, the fluid velocity is zero and the background density profile \rev{$\rho_{\text{eq}}(z)$} weakly departs from a constant reference density \rev{$\rho_0$} so that the Boussinesq approximation applies. We further assume that \rev{$\rho_{\text{eq}}(z)$} decreases linearly with $z$ such that the Brunt-V\"ais\"al\"a frequency $N = \sqrt{(-g/\rev{\rho_0})\text d\rev{\rho_{\text{eq}}}/\text dz}$ is constant. The hydrostatic pressure $p_0(z)$ is then defined by its vertical gradient $p_{0,z} \equiv \text d p_0/\text dz= -\rev{\rho_0} g$. The hydrodynamic problem 
is posed on the $f$-plane, with $f$ the Coriolis parameter. Our aim is to find perturbations of this state driven by the interaction of a background barotropic tidal flow with the bottom topography.
The barotropic flow oscillates with an angular frequency $\omega \in (f, N)$ for $N>f$ and is associated with a constant volume flux $Q$ \rev{corresponding to a uniform current $Q/h_{\pm}\cos(\omega t)$ at $x\rightarrow\pm\infty$}.
\begin{figure}
    \centering
    \centerline{\includegraphics{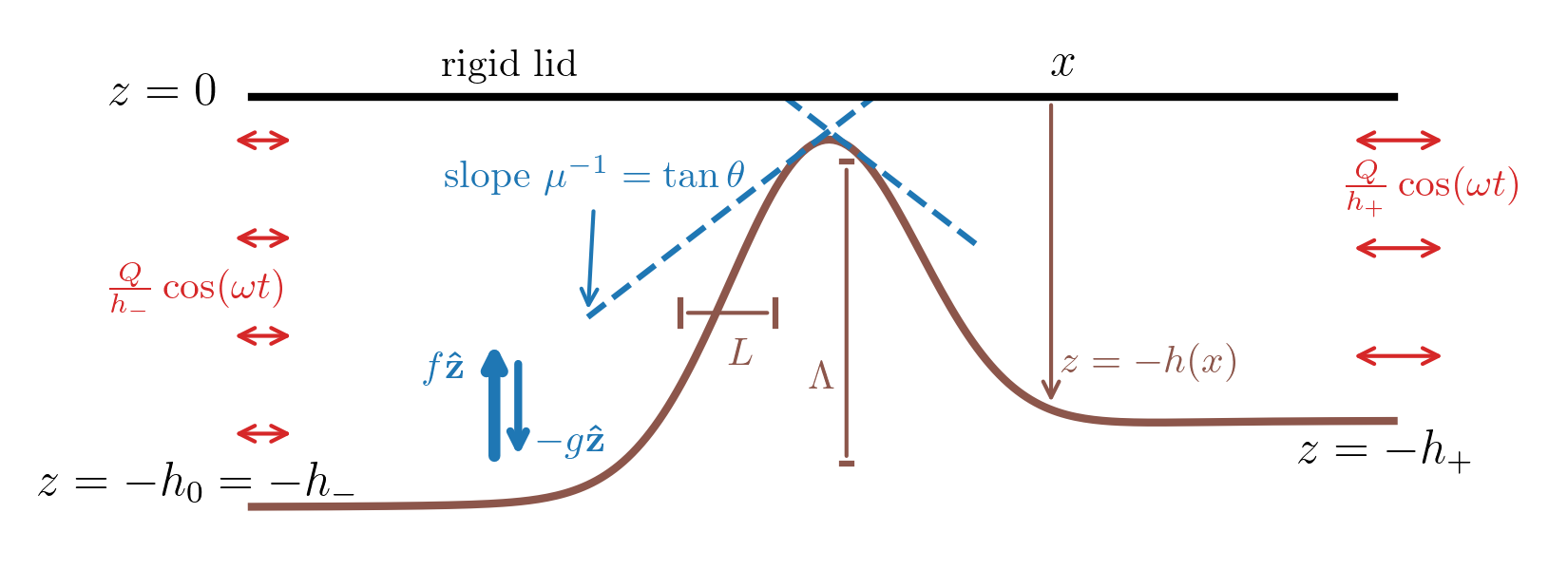}}
    \caption{Sketch of the set-up and summary of several parameters, used in this article. We use a supercritical topography ($\varepsilon>1$) for purposes of illustration, although our model can also be applied to subcritical ones.}
    \label{fig:sketch}
\end{figure}

Under the Boussinesq approximation, our setup is characterised by \rev{eight} dimensional parameters (\rev{$h_{\pm}$}, $L$, $\Lambda$, $f$, $N$, $\omega$, and $Q$), which we summarise in figure \ref{fig:sketch}, measured in combinations of meters and seconds. In the case of an isolated ridge, $h_0$ is the depth as $x\rightarrow\pm\infty$, and  $\Lambda$ \rev{is the maximum height of the ridge, $\Lambda =\max\{h_0-h\}$}. For a shelf, we assume that $ h_- > h_+$ ($h_-$ is the oceanward depth) \rev{so that the maximum height is $\Lambda=\rev{\max\{h_0-h\}}$, where we have also chosen $h_0=h_-$ as the characteristic depth.} A complete dynamical description of our system therefore requires five non-dimensional numbers. First, we introduce a ``funnelling ratio'' that measures the reduction in cross-section of the flow,
\begin{equation}
    \label{eq:delta}
    \delta = \frac{\max\{\rev{h_0-h}\}}{h_0} \sim \frac\Lambda{h_0}.
\end{equation}
The second and third parameters are the non-dimensional frequency $\omega/f$ and the characteristic slope of the internal wave,
\begin{equation}
    \mu^{-1} = \sqrt{\frac{\omega^2 - f^2}{N^2 - \omega^2}} = \tan\theta,
    \label{eq:mu}
\end{equation}
$\theta$ being the angle of the free internal wave group velocity with respect to the horizontal plane (see Eq.~\ref{eq:BVP1_freq}).
Note that internal waves are hydrostatic to a good approximation when $\omega\ll N$, or equivalently, $\mu\gg 1$.
Our fourth and fifth parameters are the relative steepness $\varepsilon$,
\begin{equation}
    \label{eq:epsilon}
    \varepsilon = \mu\max\{|\partial_x h|\} \sim \frac{\mu\Lambda}L\,,
\end{equation}
and the tidal excursion $\tau$ defined by
\begin{equation}
    \label{eq:tau}
    \tau = \frac{Q}{(h_0 - \Lambda)\omega L}.
\end{equation}
The \rev{parameter $\varepsilon$ represents the ratio $\tan \alpha/\tan \theta$ where $\alpha$ is the maximum inclination of the topography with respect to the horizontal plane. Its} 
purpose  is to measure the criticality of the topography, with $\varepsilon < 1$ ($>1$) corresponding to the subcritical (supercritical) regime. The parameter $\tau$ compares the typical displacement amplitude of a water parcel above topography, $Q/[(h_0-\Lambda)\omega]$, with the horizontal scale $L$.
If $\tau$ is finite, the curvature of the particle trajectories at the bottom generates internal waves with frequencies other than $\omega$ \citep{Bell75}.
To ensure monochromatic disturbances, we therefore assume $\tau \ll 1$.
In the ocean, for the lunar semi-diurnal tide $M_2$, the tidal excursion over flat bottom $Q/(\omega h_0)$ is $O(100\text{ m})$ \citep{Bell75}, and therefore even a moderate topographic width $L=O(10\text{ km})$ would satisfy this so-called ``acoustic limit''.

Of the five parameters defined above, $\omega/f= O(1)$ corresponds to a tidal component at mid-latitudes, and we will keep $\mu^{-1}$ fixed to a small value (for illustration purposes, $f=10^{-4}$ s$^{-1}$ is the value around latitude 45$^\circ$N, $\omega/f = 1.4$ for the M$_2$ component and we will use $N = 1.5\times 10^{-3}$ s$^{-1}$, which implies $\mu\approx 15$). Starting in Section \ref{sec:Validation}, $\varepsilon$ and $\delta$ are the parameters that we vary primarily.
Finally, 
we adopt a standard fixed value for $Q = 120$ m$^2$\,s$^{-1}$  corresponding e.g.\ to a barotropic velocity amplitude at $x\to-\infty$ of $U_0 =Q/h_0 = 4$ cm\,s$^{-1}$  and a depth $h_0 = 3$ km.
As $\varepsilon$ and $\delta$ vary, the value of $\tau$ therefore varies between calculations while remaining small.

With the parameters defined above, we can discuss the linearity of our equations.
A first way to define it is to estimate the susceptibility of radiated internal waves to undergo instabilities, which, in the $\tau\ll 1$, $\varepsilon \ll 1$ regime, is small when  $\varepsilon\tau\ll1$ \citep[e.g.,][]{Balmforth_Ierley_Young_2002, Garrett_Kunze_2007}.
In other words, the $\tau\ll 1$, $\varepsilon \ll 1$ regime is linear by construction.
However, regardless of the value of $\tau$, linearity breaks down as $\varepsilon$ increases, implying that this parameter is a better measure of linearity \citep{Garrett_Kunze_2007, buhler_muller_2007, Grisouard2012}.
In this article, we adopt the common approach of letting $\varepsilon$ be significantly supercritical in some cases while always solving a linear set of equations \citep{Petrelis_et_al_2006, griffiths2007, Balmforth_Neil_Peacock_2009, echeverri_peacock_2010, Mathur2016}. Indeed, we are primarily interested in predicting conversion from a given large-scale barotropic forcing to a topography-scale response.
The subsequent non-linear evolution of this response, which could take the form of different instabilities \citep[see][for a review]{2018_ARFM_DauxoisJOV}, could be addressed by separate parameterised procedures such as that of \citet{2009_JPO_MullerBuehler}, but would be beyond the scope of this article.

Based on the discussion above, we neglect nonlinear effects and diffusion of momentum and buoyancy, and we model the flow by the linearised, inviscid Boussinesq equations. Introducing the buoyancy $b= -g(\rho/\rev{\rho_0} - 1)$, where $\rho$ is the total density field, the governing equations are
\begin{subequations}\label{eq:Euler_eqs}
\begin{align}
    u_t - f v			& =  -p_x,\quad v_t + f u			 =  0, \label{eq:Eulx} \\
	w_t				    & =  -p_z + b, \label{eq:Eulz} \\
    b_t + N^2 w		    & =  0, \label{eq:buoy}\\
    u_x  + w_z	        & =  0, \label{eq:incompr}
	\end{align}
\end{subequations} 
where $(u,v,w)$ are the velocity components, $p$ is the associated pressure perturbation divided by \rev{$\rho_0$}, and subscripts denote partial derivatives. On the boundaries $z=-h$ and $z=0$, we have the impermeability conditions 
\refstepcounter{equation}
\[
h_x u(x,-h) + w(x,-h)  = 0\quad\text{and}\quad w(x,0)  = 0,
\eqno{(\theequation{\text{a},\text{b}})}\label{eq:BCs}
\]
and we also assume that the total
flux through a vertical cross-section is oscillating with frequency $\omega$ and constant amplitude $Q$, 
\begin{align}\label{eq:flux}
\int_{-h}^0u \dd{z} = Q \cos\omega t.
\end{align}
Introducing the time harmonic stream function $\psi = \Re\{\phi(x,z)\ee{-i\omega t}\}$, with $u = -\psi_z$, $w = \psi_x$, where $\phi$ is a complex amplitude and $\Re$ stands for the real part, Eqs. \eqref{eq:Euler_eqs}--\eqref{eq:flux} lead to a 
hyperbolic boundary value problem (BVP) on $\phi$,
\begin{align}\label{eq:BVP1_freq}
 \Lin_{\mu}\phi := \left(\partial_x^2 - \mu^{-2}\partial_z^2\right) \phi & = 0,\quad
\phi(x,0) =0\quad\text{and}\quad
\phi(x,-h) = Q,
\end{align}
where $\mu$ is defined in Eq.\ \eqref{eq:mu} \citep{Vlasenko,Garett_Gerkemma_2007}. Eq. \eqref{eq:BVP1_freq}, also known as the boundary forcing formulation, is completed by the requirement that  $\phi$ has the form of a barotropic flow plus a flow radiating away from the topography in the form of internal waves. We introduce the barotropic flow in the following subsection. 

\subsection{Barotropic flow}\label{subsec:barotropic}
The barotropic velocity components $U$, $V$, $W$ and scaled pressure $P$ satisfy Eqs.\ \eqref{eq:Euler_eqs}-\eqref{eq:BCs} where the buoyancy in the vertical momentum equation \eqref{eq:Eulz} is absent, and Eq. \eqref{eq:buoy} acts as a diagnostic equation for the induced buoyancy $B$, see Eqs. \eqref{eq:Euler_Barotrop} in appendix \rev{\ref{app:barotropic}}.
Introducing a barotropic stream function $\Psi = \Re\{\Phi \ee{-\ci \omega t}\}$ for some time independent function $\Phi$ with $U = -\Psi_z$ and $ W = \Psi_x$, we obtain from \eqref{eq:Euler_Barotrop},
\begin{align}\label{eq:BarotropicBVP}
\Phi_{xx}+ \mu_0^{-2}\Phi_{zz}=0, \quad \Phi(x,0)=0\quad\text{and}\quad  \Phi(x,-h)=Q,
\end{align}
with $\mu_0^{-2}=1 - f^2/\omega^2$ \citep{Garett_Gerkemma_2007} where the boundary conditions ensure that the total mass transport through any vertical
cross-section is solely due to the barotropic flow. Note that the PDE in \eqref{eq:BarotropicBVP} is elliptic and is also obtained by \eqref{eq:BVP1_freq} with $N\equiv 0$.
The solution of \eqref{eq:BarotropicBVP} can be written as 
\begin{align}\label{eq:barotropic_sol}
    \Phi = \Phi^{(0)}+\Phi^{\text{r}},\quad\text{with}\quad  \Phi^{(0)} = -Qz/h,
\end{align}
where $\Phi^{(0)}$ represents the hydrostatic part of the barotropic flow. It is obtained by neglecting $W_t$ in \eqref{eq:Eulz_Barotrop} and, consequently, $\Phi_{xx}$ in \eqref{eq:BarotropicBVP}; see Appendix \ref{app:barotropic} for a detailed derivation.  $\Phi^{\text{r}}$ represents the residual non-hydrostatic part which solves 
\begin{align}\label{eq:BarotropicBVP_NH}
\Phi^{\text{r}}_{xx}+ \mu_0^{-2}\Phi^{\text{r}}_{zz}=-\Phi^{(0)}_{xx}, \quad \Phi^{\text{r}}(x,0)=0\quad\text{and}\quad \Phi^{\text{r}}(x,-h)=0.
\end{align}
Note that $\Phi$ is spatially non-uniform, that is, it depends on $z$ and $h(x)$. Also note that $\Phi^{\text{r}}$ vanishes at $x\rightarrow\pm\infty$. We obtain a general semi-analytical solution of \eqref{eq:BarotropicBVP_NH} that is valid for arbitrary smooth $h$ by means of a modal decomposition (subsection \ref{subsec:CMS}). In Appendix \ref{app:barotropic}, we derive a perturbative solution that is valid if $h_0/L\ll 1$.

In fact, $\Phi^{(0)}$ coincides with the leading order \rev{part} of this solution. \rev{The corresponding velocities are given by $[U^{(0)},W^{(0)}]=[Q/h, Qz(1/h)_x]\cos(\omega t)$. As $x\rightarrow\pm\infty$, $W^{(0)}=0$ and $U^{(0)}=Q /h_{\pm}\cos(\omega t)$ coincide with the spatially uniform (depth-averaged) barotropic currents far from the topography.} $U^{(0)}$ is also used as a forcing in the IT model of \cite{griffiths2007} in the case of a shelf without coastline.

\subsection{The internal tide generation problem\label{sec:ITgeneration}}
Introducing 
$\phi^{\#}=\phi-\Phi$,
we obtain from \eqref{eq:BVP1_freq} the BVP
\begin{align}\label{eq:BVP_freq0NH}
\Lin_{\mu} \phi^{\#}  = -\Lin_{\mu}\Phi,\quad
\phi^{\#}(x,0)=0\quad\text{and}\quad
\phi^{\#}(x,-h)= 0,  
\end{align}
which shows that the barotropic flow $\Phi$ forces a purely baroclinic response $\phi^{\#}$. Alternatively, introducing $\phi^{\dagger}=\phi^{\#} +\Phi^{\text{r}}$ and exploiting the linearity of $\Lin_\mu$, the BVP \eqref{eq:BVP_freq0NH} becomes
\begin{align}
\label{eq:BVP_freq0H}
\Lin_{\mu} \phi^{\dagger}  = -\Lin_{\mu}\Phi^{(0)} =-\Phi^{(0)}_{xx},\quad
\phi^{\dagger}(x,0)=0\quad\text{and}\quad
\phi^{\dagger}(x,-h)= 0,
\end{align}
which shows that the hydrostatic part of barotropic flow, $\Phi^{(0)}$, forces a baroclinic response plus a non-hydrostatic barotropic one \citep{Garett_Gerkemma_2007}. This formulation is referred to as the body-forcing formulation since the forcing appears only in the wave equation \citep{Garett_Gerkemma_2007,Garrett_Kunze_2007}. An advantage of working with such a formulation is that the unknown field  
satisfies homogeneous Dirichlet conditions that make the application of a coupled-mode approach straightforward (Section \ref{sec:CMS}). Here, we shall proceed with \eqref{eq:BVP_freq0H} mainly because $\Phi^{(0)}$ is given by the simple explicit expression \eqref{eq:barotropic_sol}; $\Phi^{\text r}$ can be computed independently in order to extract the purely baroclinic field  $\phi^{\#}=\phi^{\dagger}-\Phi^{\text{r}}$ if needed. Also, $\Phi^{\text r}$ \rev{results in a spatially trapped correction to the flow}; namely, it vanishes as $x\rightarrow\pm\infty$ and thus does not influence the far-field energy.
Eqs.\! \eqref{eq:BVP_freq0H} are supplemented with radiation conditions ensuring that waves generated in the interior of the domain propagate outward as plane waves. Thus, 
we have
\begin{align}\label{eq:radiation_conditions}
\rev{\phi^{\dagger} = \sum\limits_{n=1}^{\infty}c^\pm_n\ee{\pm \ci k^{\pm}_n x}\sin\left(\frac{n\pi z}{h_{\pm}}\right),\ \ \text{with}\ \ k^{\pm}_n = \frac{n\pi}{\mu h_{\pm}},\ c^{\pm}_n\in\mathbb{C}, \ \ \text{as}\ \ x\rightarrow\pm\infty}.
\end{align}
It is useful to write down expressions for the flow fields in terms of the amplitudes of the stream functions. From the above analysis, $\phi = \Phi^{(0)}+\Phi^{\text{r}}+\phi^{\#} =\Phi^{(0)} + \phi^{\dagger}$ and we may introduce the corresponding definitions
\begin{align}\label{eq:fdecomposition}
\begin{aligned}
    \xi &= \Xi^{(0)} + \Xi^{\text r} + \xi^\# = \Xi^{(0)} + \xi^\dagger,
\end{aligned}
\end{align}
where $\xi$ is a placeholder for any of $u$, $v$, $w$, $b$ or $p$, and $\Xi$ is a placeholder for any of  $U$, $V$, $W$, $B$ or $P$. \rev{The flow components appearing in Eq. (2.15) satisfy the relations}
\begin{equation}\label{eq:Relations}
\begin{pmatrix}
U^{\diamond}& V^{\diamond}& W^{\diamond}&  B^{\diamond}\\
u^\star&v^\star& w^\star& b^\star
\end{pmatrix}
=  \Re\left\{
\begin{pmatrix}
-\Phi^{\diamond}_z & \frac{\ci f}\omega\Phi^{\diamond}_z & \Phi^{\diamond}_x & -\frac{\ci N^2}{\omega}\Phi^{\diamond}_x\\
-\phi^\star_z & \frac{\ci f}\omega\phi_z^\star & \phi_x^\star & -\frac{\ci N^2}{\omega}\phi^\star_x
\end{pmatrix}\ee{-\ci \omega t}\right\},
\end{equation}
where the superscript $\star$ stands for either $\dagger$ or $\#$\rev{,} and the superscript $\diamond$ stands for either $(0)$ or $\text{r}$. 
The relations for $v^{\star},V^{\diamond}$  (resp. $b^{\star},B^{\diamond}$) follow from the second equation in \eqref{eq:Eulx} (resp. \eqref{eq:buoy}) and the assumption that all fields have the same time periodicity. A similar expression can be derived for the pressure containing additionally boundary terms.
For easy reference, we summarise the notations for the different flow fields in table~\ref{tab:fields_notations}. Plugging the second equality of \eqref{eq:fdecomposition} into \eqref{eq:Euler_eqs}--\eqref{eq:BCs} and taking into account \eqref{eq:Euler_Barotrop_asympt}--\eqref{eq:BCs_Barotrop_asympt} with $i=0$, we obtain
\begin{subequations}\label{eq:Euler_dagg}
\begin{align}
    u^{\dagger}_t - f v^{\dagger}			 & =  -p^{\dagger}_x, \quad
	v^{\dagger}_t + f u^{\dagger}			 =  0,\label{eq:Eul_Baroclin} \\
	w^{\dagger}_t				    & = -p^{\dagger}_z + b^{\dagger} + \left(1-\frac{\omega^2}{N^2}\right)B^{(0)}, \label{eq:Eulz_Baroclin}
	\end{align}
\end{subequations}
\begingroup\abovedisplayskip=0pt
\begin{align}	
b^{\dagger}_t + N^2 w^{\dagger}		& = 0, \label{eq:buoy_Baroclin}
\end{align}
\endgroup
\begingroup\abovedisplayskip=-5pt
\begin{align} 
 u^{\dagger}_x  + w^{\dagger}_z	& =  0,\label{eq:incompr_Baroclin}
\end{align}
\begin{align}
h_x u^{\dagger}(x,-h)+w^{\dagger}(x,-h) & = 0,\quad w^{\dagger}(x,0)  = 0.
\label{eq:BCs_Baroclin}
\end{align}
\endgroup
In deriving \eqref{eq:Eulz_Baroclin}, we used the relation $B^{(0)} = N^2 W^{(0)}_t/ \omega^2$, which itself derives from \eqref{eq:Relations}. In terms of $\xi^{\#}$, the above equations stay the same except for \eqref{eq:Eulz_Baroclin}, which becomes 
\begin{align}
w^{\#}_t				     = -p^{\#}_z + b^{\#} + B, \label{eq:Eulz_Baroclin_diese}    
\end{align} 
showing that the baroclinic flow is forced by the buoyancy force 
created by the barotropic flow \citep{Garett_Gerkemma_2007}.
\begin{table}
    \centering
    \begin{tabular}{c|c|c}
        Notation & Field type \& Equations & Related definitions \\\hline 
        $\phi$, $\xi$ & Total, \eqref{eq:BVP1_freq},  \eqref{eq:Euler_eqs}--\eqref{eq:BCs} & $\xi = \xi^\#+\Xi$ \\
        $\Phi$, $\Xi$ & Barotropic, \eqref{eq:BarotropicBVP},  \eqref{eq:Euler_Barotrop}--\eqref{eq:BCs_Barotrop} & $\Xi = \Xi^{(0)}+\Xi^{\text{r}}$\\
        $\Phi^{(0)}$, $\Xi^{(0)}$ & Hydrostatic barotropic, \eqref{eq:barotropic_sol}, \eqref{eq:Euler_Barotrop_asympt}--\eqref{eq:BCs_Barotrop_asympt}  & \\
         $\Phi^\text{r}$, $\Xi^\text{r}$ & Non-hydrostatic barotropic, \eqref{eq:BarotropicBVP_NH} & $\Xi^\text{r} = \Xi - \Xi^{(0)}$\\
        $\phi^\dagger$, $\xi^\dagger$ &  Baroclinic-barotropic(non-hydrostatic), \eqref{eq:BVP_freq0NH}, \eqref{eq:Euler_dagg}--\eqref{eq:BCs_Baroclin}   & $\xi^\dagger = \xi^\# + \Xi^\text{r}$  \\
        $\phi^{\#}$, $\xi^{\#}$ & Baroclinic, \eqref{eq:BVP_freq0NH}, \eqref{eq:Eul_Baroclin}, \eqref{eq:Eulz_Baroclin_diese},  \eqref{eq:buoy_Baroclin}--\eqref{eq:BCs_Baroclin}  
    \end{tabular}
    \caption{The notations for the fields we use.
    The symbol $\xi$ stands for $u$, $v$, $w$, $b$, $p$, and $\Xi$ for their capitalized versions. The relations in the third column hold for $\xi$ and $\Xi$ replaced by $\phi$ and $\Phi$.}
    \label{tab:fields_notations}
\end{table}

\subsection{Energy equation and conversion rate}
We derive here the energy equation for the above IT generation problem and use it to define the energy conversion rates. 

The dot product of \eqref{eq:Euler_dagg}--\eqref{eq:buoy_Baroclin} with  $\left(\mathbf{u}^{\dagger},b^{\dagger}/N^2\right) \equiv (u^{\dagger}, v^{\dagger}, w^{\dagger},b^{\dagger}/N^2)$ gives
\begin{align}
    \mathcal{E}^\dagger_t + \boldsymbol{\nabla\cdot}\left(p^\dagger\mathbf{u}^\dagger\right) = \left(1-\frac{\omega^2}{N^2}\right)B^{(0)} w^\dagger,\quad \text{with}\quad \mathcal{E}^\dagger = \frac{1}{2}(\mathbf{u}^{\dagger})^2 +\frac{1}{2}\frac{(b^{\dagger})^2}{N^2},\label{eq:en_density}
\end{align}
where we have used \eqref{eq:incompr_Baroclin} and where $\boldsymbol\nabla \equiv \left(\partial_x, 0, \partial_z \right)$.
Integrating \eqref{eq:en_density} over the domain $\Omega = [-\infty,+\infty]\times [-h,0]$, 
 using the divergence theorem and \eqref{eq:BCs_Baroclin}, we obtain 
\begin{align}
    \left(\int_\Omega \mathcal{E}^\dagger \dd{\Omega}\right)_t + \left[\int_{-h}^0 p^{\dagger} u^{\dagger} \dd{z}\right]^{+\infty}_{-\infty} 
    =  \left(1-\frac{\omega^2}{N^2}\right)\int_{\Omega}B^{(0)} w^{\dagger} \dd{\Omega},
\end{align}
with $[\,\cdot\,]^{+\infty}_{-\infty}=\lim_{x\to\infty}(\cdot) - \lim_{x\to-\infty}(\cdot)$. 

Applying the time average operator $\langle \,\cdot\,\rangle  = 1/T \int_0^T \cdot\, \dd{t}$, with $T = 2\pi/\omega$, and taking into account the periodicity of $\mathcal{E}^{\dagger}(t)$, we find
\begin{align}\label{eq:energy_eq_avg}
\begin{aligned}C_{+} -C_{-}\stackrel{\text{def}}{=}\left[\int_{-h}^0 \langle p^{\dagger} u^{\dagger}\rangle \dd{z}\right]^{+\infty}_{-\infty}&=
    \left(1-\frac{\omega^2}{N^2}\right)\int_{\Omega} \langle B^{(0)} w^{\dagger}\rangle  \dd{\Omega} \stackrel{\text{def}}{=}C_{\text{int}},\end{aligned}
\end{align}
where we have defined the energy conversion rates $C_{\pm}$ that represent the rates at which energy is radiated at $\pm\infty$, and the total energy convergence rate $C_{\text{int}}$ given as a volume integral. 
Note that the non-hydrostatic barotropic flow does not contribute in \eqref{eq:energy_eq_avg} because $u^{\dagger}=u^{\#}+U^r \rightarrow u^{\#} $ as $\rev{x\rightarrow \pm\infty}$ and 
$\langle B^{(0)} w^{\dagger}\rangle = \langle B^{(0)} w^{\#}\rangle + \langle B^{(0)} W^r\rangle = \langle B^{(0)} w^{\#}\rangle$ since $B^{(0)}$ and $W^r$ are out of phase by $\pi/2$ (Eq.~\eqref{eq:Relations}). Thus, \eqref{eq:energy_eq_avg} remains valid with $\dagger$ replaced by $\#$.
\rev{We proceed by expressing $C_{\pm}$ and $C_{\text{int}}$ in terms of $\phi^{\dagger}$ and $\Phi^{(0)}$.}
\rev{Using} integration by parts and the fact that $\psi^{\dagger}(x,0) =\psi^{\dagger}(x,-h)=0$ \rev{we} obtain $\int_{-h}^0 \left\langle  p^{\dagger} u^{\dagger}\right\rangle  dz  =  \int_{-h}^0 \left\langle p_z^{\dagger} \psi^{\dagger}\right\rangle  dz$. Then,
expressing $p_z^{\dagger}$ from \eqref{eq:Eulz_Baroclin} using \eqref{eq:Relations}, we find
\begin{align}\label{eq:pzpsi}
\langle p_z^{\dagger}\psi^{\dagger}\rangle  = \frac{\omega^2-N^2}{\omega} \langle\Re\{\ci\phi^{\dagger}_x \ee{-\ci\omega t}\}\Re\{\phi^{\dagger} \ee{-\ci\omega t}\}\rangle\quad \text{as}\quad x\rightarrow\pm\infty.  
\end{align}
Writing $\ee{-\ci\omega t} = \cos\omega t - \ci \sin\omega t $ and noting that $\langle\cos\omega t\sin\omega t\rangle = 0$, $\langle\cos^2\omega t\rangle = \langle\sin^2\omega t\rangle = 1/2$ we obtain
\begin{align}\label{eq:Conv}
C_{\pm}  = \rev{\frac{\omega^2-N^2}{2\omega}}\int_{-h}^0 \Im\{\phi^{\dagger}\overline{\phi_x^{\dagger}}\}  \dd{z}, \quad \text{as}\quad x\rightarrow\pm\infty, 
\end{align}
where the overline denotes the complex conjugate and $\Im$ the imaginary part. Similarly,
\begin{align}\label{eq:ConvInt}
    C_{\text{int}} &=  \left(1-\frac{\omega^2}{N^2}\right)\frac{N^2}{2\omega}\int_{\Omega} \Phi^{(0)}_x\Im\{\,\overline{\phi^{\dagger}_x}\,\} \dd{\Omega}.
\end{align}
\rev{Using the radiation conditions \eqref{eq:radiation_conditions} in Eq.\ \eqref{eq:Conv}, we see that $C_+\geq 0$ (resp. $C_-\leq 0$) as $x\rightarrow +\infty$ (resp. $x\rightarrow -\infty$). Eqs.\ \eqref{eq:Conv} and \eqref{eq:ConvInt} provide us with  two ways of calculating the total conversion rate, either by using the the far-field baroclinic flow or by using a barotropic-baroclinic interaction term defined in the entire fluid domain. This fact will be used in Section \ref{sec:Validation} for validation purposes.}


\section{Modal decomposition}\label{sec:CMS}
\subsection{Stream function modal representation}\label{sec:representation}
We reformulate the IT generation problem \eqref{eq:BVP_freq0H}--\eqref{eq:radiation_conditions} by representing $\phi^{\dagger}$ as
\begin{align}
\phi^{\dagger}(x,z) = \sum\limits_{n=1}^{\infty}\phi_n(x)Z_n(z;x),\label{eq:RL_representation}
\end{align}
where $\{Z_n(z;x)\}_{n=0}^{\infty}$ are prescribed vertical basis functions with a parametric dependence on $x$ and $\{\phi_n(x)\}_{n=0}^{\infty}$ are unknown complex modal amplitudes to be determined. For the expansion \eqref{eq:RL_representation} to be exact, the set $\{Z_n(z;x)\}_{n=0}^{\infty}$ must be complete. \rev{In the present constant stratification case}, this set is obtained as the set of eigenfunctions of a Sturm-Liouville problem parameterised by $x$, also called the ``reference waveguide'' \citep{Brekhovskikh},

\begin{align} \label{eq:SLproblem}
Z_{n,zz} + \rev{\frac{n^2\pi^2}{h^2}}  Z_n = 0,\quad
        Z_n(0 ;x) = 0,\quad
        Z_n(-h ;x)= 0.
\end{align}

\rev{The eigenfunctions $Z_n$ are given by}

\begin{align}
 \rev{Z_n(z;x) =  \sin\left(\frac{n\pi z}{h}\right),}\label{eq:RL_vertical_function}   
\end{align}
 \rev{and satisfy} the orthogonality relation $\int_{-h}^{0}Z_n Z_m dz = h\delta_{nm}/2$, where $\delta_{nm}$ is the Kronecker delta. Note that \eqref{eq:RL_representation} satisfies exactly and term by term the boundary conditions in \eqref{eq:BVP_freq0H}. It follows from \eqref{eq:RL_representation} and \eqref{eq:RL_vertical_function} that the $\phi_n$'s are defined by
\begin{align}\label{eq:def_mod}
    \phi_n = \frac{2}{h}\int_{-h}^0\phi^{\dagger} Z_n dz =\rev{\frac{2h^2}{\pi^3 n^3}}\left( \left[\phi^{\dagger}_{zz}Y_n\right]_{-h}^0 - \int_{-h}^0 \phi^{\dagger}_{z z z}Y_n \dd{z}\right),
\end{align}
 where $Y_n=\cos(\rev{n\pi z / h})$ and the second equality is obtained after three integrations by parts, which are allowed provided $\phi^{\dagger}$ is sufficiently smooth, and use of the boundary conditions in \eqref{eq:BVP_freq0H}. 
This shows that $\|\phi_n\|_{\infty}:=\max |\phi_n|=O(n^{-3})$ and that \eqref{eq:RL_representation}  
converges uniformly in this case.
Similar estimates are obtained for $\|\phi_{n,x}\|_{\infty}$ and $\|\phi_{n,xx}\|_{\infty}$ by adapting the procedure developed in \citet[Section 4]{Ath_Pap_2017} and suffice to establish the term-wise differentiability of \eqref{eq:RL_representation} required for the exact modal reformulation of \eqref{eq:BVP_freq0H}. 

\citet{griffiths2007} use a similar expansion for $u$ and derive a series representation for $w$ by using the incompressibility and the bottom-boundary conditions. The maximum decay rate in this case is $O(n^{-2})$ due to the non-vanishing of $u$ on the boundaries. Note also that in contrast to \eqref{eq:RL_representation}, the truncated version of this expansion does not satisfy term by term the bottom boundary condition.  On the other hand,  \cite{Kelly2016} uses two sets of eigenfunctions: one for $u$ and $p$, and the other for $w$ and $b$. In this approach, $w$ vanishes identically on the bottom which is not the case for arbitrary $h$. Consequently, this approach should be regarded as an approximation, see the discussion in \cite{Kelly2016b} for more details. Despite \eqref{eq:RL_representation} being limited to 2D flows, its major advantage is that it satisfies the bottom boundary condition in \eqref{eq:BVP_freq0H} exactly and term by term, and exhibits faster convergence in comparison with existing approaches.

\subsection{The Coupled-Mode System}\label{subsec:CMS}
We proceed by projecting Eqs.\! \eqref{eq:BVP_freq0H}--\eqref{eq:radiation_conditions} onto \eqref{eq:RL_vertical_function}. Substituting \eqref{eq:RL_representation} into \eqref{eq:BVP_freq0H}, multiplying with $Z_n$ and integrating over the interval $[-h(x),0]$, we find that  $\{\phi_n(x)\}_{n=1}^{\infty}$ solves the following Coupled-Mode System (CMS), for $m\geq 1$, 
\begin{align}\label{eq:CMS}
\phi_{m,xx}+\rev{\frac{m^2\pi^2}{\mu^2h^2}}\phi_m+\sum\limits_{n=1}^{\infty}\left[\frac{b_{mn}h_x}{h}\phi_{n,x} + \left(\frac{c_{mn}h_x^2}{h^2}+\frac{d_{mn}h_{xx}}{h}\right)\phi_n\right]
= 2 g_m h\left(\frac{1}{h}\right)_{xx},
\end{align}
 where $b_{mn}$, $c_{mn}$, $d_{mn}$ are $x$-independent coefficients given in Appendix \ref{app:ABC}, and $g_m =  Q (-1)^{\rev{m+1}}/(m\pi)$. 
Substituting \rev{Eq.}\ \eqref{eq:RL_representation} into \eqref{eq:radiation_conditions}, we obtain
\begin{align}\label{eq:BVP2_RCleft_m}
\sum\limits_{n=1}^{\infty}\phi_n(x)Z_n(x;z) = \sum\limits_{n=1}^{\infty}c_n^\pm\exp(\pm \ci k^{\pm}_n x)\sin\left(\rev{\frac{n\pi z}{h_{\pm}}}\right)\quad\text{as}\quad x\rightarrow \pm\infty.
\end{align}
Multiplying both sides by $Z_q$, $q\geq 1$, integrating over $[-h,0]$ and taking into account that  $Z_n\rightarrow \sin(\rev{n\pi z/ h_{\pm}})$ as $ x\rightarrow \pm \infty$, we obtain $\phi_n = c_n^\pm\exp(\pm \ci k^{\pm}_n x)$,
and therefore
\begin{align}\label{eq:rad_cond_left}
\phi_{n,x} \rev{\mp} \ci k^{\pm}_n\phi_n= 0,\quad\text{as}\quad x\rightarrow {\pm}\infty.
\end{align}
Thus, Eqs.\! \eqref{eq:BVP_freq0H}--\eqref{eq:radiation_conditions} are exactly reformulated as the CMS  \eqref{eq:CMS}--\eqref{eq:rad_cond_left}. Solving the latter, we may reconstruct the solution of the former by using \eqref{eq:RL_representation}. The energy conversion rate\rev{s} $C_{\pm}$ \rev{are} evaluated by \eqref{eq:Conv}: 
\rev{\begin{align}
    C_{\pm} = \frac{\omega^2-N^2}{2\omega}\frac{h}{2} \sum\limits_{n=1}^{\infty}\Im 
    \{\phi_n^{\dagger}\overline{\phi^{\dagger}}_{\!\!\!n,x}\} = \pm \frac{N^2-\omega^2}{4\omega\mu}\pi \sum\limits_{n=1}^{\infty}n \phi_n^{\dagger}\overline{\phi^{\dagger}}_{\!\!\!n}\quad\text{as}\quad x\rightarrow {\pm}\infty,
\end{align}}
\rev{where Eqs.\ \eqref{eq:rad_cond_left} and the definition of $k_n^{\pm}$ in \eqref{eq:radiation_conditions} have been used to obtain the second equality.} The purely baroclinic field $\phi^{\#}$ is computed using $\phi^{\#} = \phi^{\dagger}-\Phi^{\text{r}}$. For the computation of $\Phi^{\text{r}}$, we apply the same modal decomposition to \eqref{eq:BarotropicBVP_NH}. The resulting CMS is given by \eqref{eq:CMS} with $\mu^2$ replaced by $-\mu_0^2$ and vanishing conditions at infinity, instead of \eqref{eq:rad_cond_left}; if $h_0/L\ll 1$, one could use instead the approximate asymptotic expression in \eqref{eq:Phi_orders_dim}.
In the following subsection, we derive a perturbative solution of the CMS \eqref{eq:CMS}--\eqref{eq:rad_cond_left} for infinitesimal topography. For arbitrary topographies, we solve the CMS numerically (see subsection \ref{subsec:numerical_scheme}).

\begin{remark}\label{rem:HA}
The body-forcing formulation \eqref{eq:BVP_freq0NH} and its coupled-mode reformulation \eqref{eq:CMS} are valid for non-hydrostatic conditions. If the hydrostatic approximation (HA) is invoked for both the baroclinic and the barotropic flow ($\omega^2\ll N^2$ and $h_0^2\ll L^2$), Eq.\ \eqref{eq:BVP_freq0NH} becomes \eqref{eq:BVP_freq0H} with $\mu^2$ replaced $\mu'^2=(\omega^2-f^2)/N^2$ and $\phi^{\dagger}$ is interpreted as a purely baroclinic response, that is, $\Phi^r=O(h^2/L^2)$ may be neglected (Appendix \ref{app:barotropic}) \citep{Garett_Gerkemma_2007}. The CMS \eqref{eq:CMS} changes accordingly. Under this assumption and using \eqref{eq:Relations},  Eq.\ \eqref{eq:energy_eq_avg} reduces to 
the energy equation derived in \cite{Gerkema2004} starting \rev{from} the hydrostatic governing equations. If instead we assume $\omega^2\ll N^2$ and $h_0^2/L^2= O(1)$ then $\Phi^r$ is not negligible but does not affect the energy flux at infinity.   In the reverse situation, $\omega^2< N^2$ and $h_0^2/L^2\ll 1$, Eqs.\ \eqref{eq:BVP_freq0NH} and \eqref{eq:CMS} hold as is but once again $\phi^{\dagger}$ is interpreted as a purely baroclinic response. For more details on the relevance of the HA we refer to the discussion in \cite{Garett_Gerkemma_2007}. Unless otherwise stated, we do not invoke the HA because the obtained mathematical simplification is minute and the precise quantification of the induced error for different values of $\mu$ is out of the scope of this work.  
\end{remark}

\subsection{Infinitesimal topography solution }
Let  $h(x) = h_0 -\epsilon\,r(x)$ with $|\epsilon|\ll 1$ and $r=O(1)$, for some characteristic depth $h_0$. Introducing the asymptotic expansion $\phi_m =\sum_{i=0}^K\epsilon^i\phi_m^{(i)}$ and Taylor-expanding $1/h$ in \eqref{eq:CMS}, we deduce that $\phi_m^{(0)}=0$ and that $\phi_m^{(1)}$ solves 
\begin{align}\label{eq:CMS_WTA}
\phi^{(1)}_{m,xx}+\ell_m^2\phi^{(1)}_m
= 2 g_m \frac{r_{xx}}{h_0},\quad\text{for}\quad m\geq 1,
\end{align}
with $\ell_m=\rev{m}\pi/(h_0\mu)$, and the radiation conditions in \eqref{eq:rad_cond_left} with $k_m^+ = k_m^- =\ell_m$. Substituting the solution \eqref{eq:phin_WTA} in \eqref{eq:Conv} we obtain the conversion rate for weak topography,
\begin{align}\label{eq:ConversionStL}
C^{\text{WTA}} = \frac{F_0}{h_0^{2}}\sum\limits_{n=1}^{\infty} |\hat{r}(\ell_n)|^2 \frac{n\pi^2}{(\mu h_0)^2}\,\,\,\,\,\text{with}\,\,\,\,\, F_0 = \frac{\left[(N^2-\omega^2)(\omega^2-f^2)\right]^{1/2}}{2\pi\omega}U\rev{_0}^2h_0^2,
\end{align}
where $\hat{r}(\xi) = \int_{-\infty}^{+\infty}\exp(-\ci x\xi)r(s) ds$ is the Fourier transform of $r$ (Appendix \ref{app:WTA}). This is the formula given in \cite{StLaurent2003} (up to multiplication with \rev{$\rho_0$}), which was first derived in 
\cite{Llewellyn2002} in the case of hydrostatic internal waves. \cite{Khatiwala2003} derived a different formula starting from a multi-frequency representation of the response fields in terms of Bessel functions. In the acoustic limit (recall discussion following Eq.\ \ref{eq:tau}), the dominant contribution to the far-field energy comes from the fundamental frequency, \rev{say,} $\omega_0$ \citep{Bell75}. The quantity $J_1(U_0 k_{j1}/\omega_0)$ appearing in \cite{Khatiwala2003}'s Eq.\ (28), where $J_1$ is the order-one Bessel function of the first kind, and $U_0$ and $k_{j1}$ are his barotropic tidal amplitude and wavenumber for the $j^\text{th}$ mode of the fundamental frequency, respectively, may be replaced by its asymptotic value $U_0k_{j1}/(2\rev{\omega_0})$ leading to our Eq.~\eqref{eq:ConversionStL} \rev{with $\omega=\omega_0$}.

\subsection{Numerical solution for arbitrary topography}\label{subsec:numerical_scheme}
We truncate the infinite CMS \eqref{eq:CMS}--\eqref{eq:rad_cond_left} by keeping the first $M$ equations and replacing the infinite domain by a finite interval $X=[x_L,x_R]$ of length $L_X = x_R-x_L$. 
We discretise $X$  with a uniform spacing $\delta x$, $\{x_i$, $i=1,N_X\}$ and we approximate the derivatives using $4^{th}$-order finite differences up to the boundary points $x=x_L,x_R$.
The corresponding formulae can be found in \citet[Appendix C]{PAPOUTSELLIS2019103579}. We thus obtain a sparse square linear system of dimension $(N_X M)^2$ for the grid values of each modal amplitude $\phi_n(x_i)$, $i=1,\dots,N_X$, $n=1,\dots,M$ which we solve by means of a $LU$ decomposition.
Using this solution, we reconstruct the field $\phi^{\dagger}$ by means of a truncated version of \eqref{eq:RL_representation}. We then compute the baroclinic fields and conversion rates using Eqs.\ \eqref{eq:Relations} and \eqref{eq:Conv}.  
We also provide an estimation of the free-surface elevation due to the IT motion given by $\eta = [p^{\#}]_{z=0}/ g$, where $[p^{\#}]_{z=0}$ is the pressure induced on $z=0$ by the baroclinic flow (Appendix \ref{app:free_surface}).

\section{Convergence, accuracy and singularity formation \label{sec:Validation}}
In this section, we introduce the topographic profiles we use in our calculations and present results showing the good performance of our semi-analytical solution. 

\subsection{Topographic profiles}
\begin{figure}
\centering
  \includegraphics[scale = 0.7]{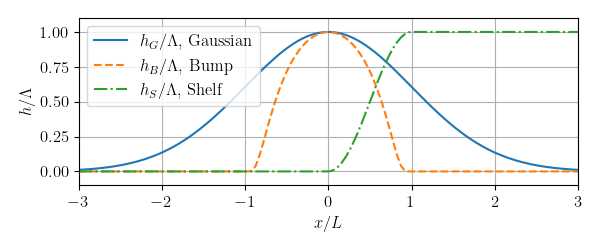} 
\caption{The three topographic profiles we consider (in the Shelf case, we have used $h_- = 0$, $h_+ = \Lambda$).}
\label{Fig:ridges}
\end{figure}

We consider two ridge profiles, namely the ``Gaussian" $h=h_0-h_\text{G}$ and the ``bump",  $h=h_0-h_\text{B}$, where 
\begin{align}\label{eq:topos}
    h_{\text{G}}=\Lambda\exp\left(-\frac{x^2}{2L^2}\right)\quad\text{and}\quad
    h_{\text{B}}=\Lambda\exp\left(1-\frac{1}{1-x^2/L^2}\right)\mathds{1}_{(-L,L)},
\end{align}
with $\mathds{1}_{(-L,L)}=1\,\,\text{in}\,\,(-L,L)$ and $\mathds{1}_{(-L,L)}=0$ otherwise. Note that the ``bump" profile has a compact support, in contrast with the Gaussian, and all its derivatives are continuous at $x = \pm L$; see figure \ref{Fig:ridges}. We also consider the case of a shelf connecting two different constant depths. The shelf profile is the same as in \citet[Section 5]{griffiths2007}, namely, \rev{$h=h_{\text{S}}$ with}
\begin{align}\label{eq:shelf_topo}
    h_\text{S} = \left\{\begin{aligned}
        & h_-\quad\text{for}\quad x\leq 0,\\
        & h_- + (h_+-h_-) \sin^2(\pi x/2L) \quad\text{for}\quad 0\leq x\leq L, \quad \text{and}\\
        & h_+\quad\text{for}\quad x\geq L.
    \end{aligned}\right.
\end{align} 
We recall from subsection \ref{subsec:posing} that two important parameters will be considered, namely, the relative topography height $\delta$ (Eq. \ref{eq:delta}) and the criticality $\varepsilon$ (Eq. \ref{eq:epsilon}) of the bottom slope.
We also recall that, for illustration purposes, $(\omega,U_0)$ corresponds to a typical $M_2$ tide, i.e., $(\omega,U_0) = (1.4\times10^{-4}\ \text{s}^{-1}, 0.04\,  \text{m\,s}^{-1})$ and that the Brunt-V\"ais\"{a}l\"{a} and Coriolis frequencies are  $ N = 1.5\times 10^{-3} \ \text{s}^{-1}$ and $f = 10^{-4} \ \text{s}^{-1}$, respectively. For all ridges, the depth far from the topography is $h_0 = 3000$~m.

\subsection{Results}
\label{sec_convergence}
\begin{figure}
\centering
\centerline{
\includegraphics[width = 14cm]{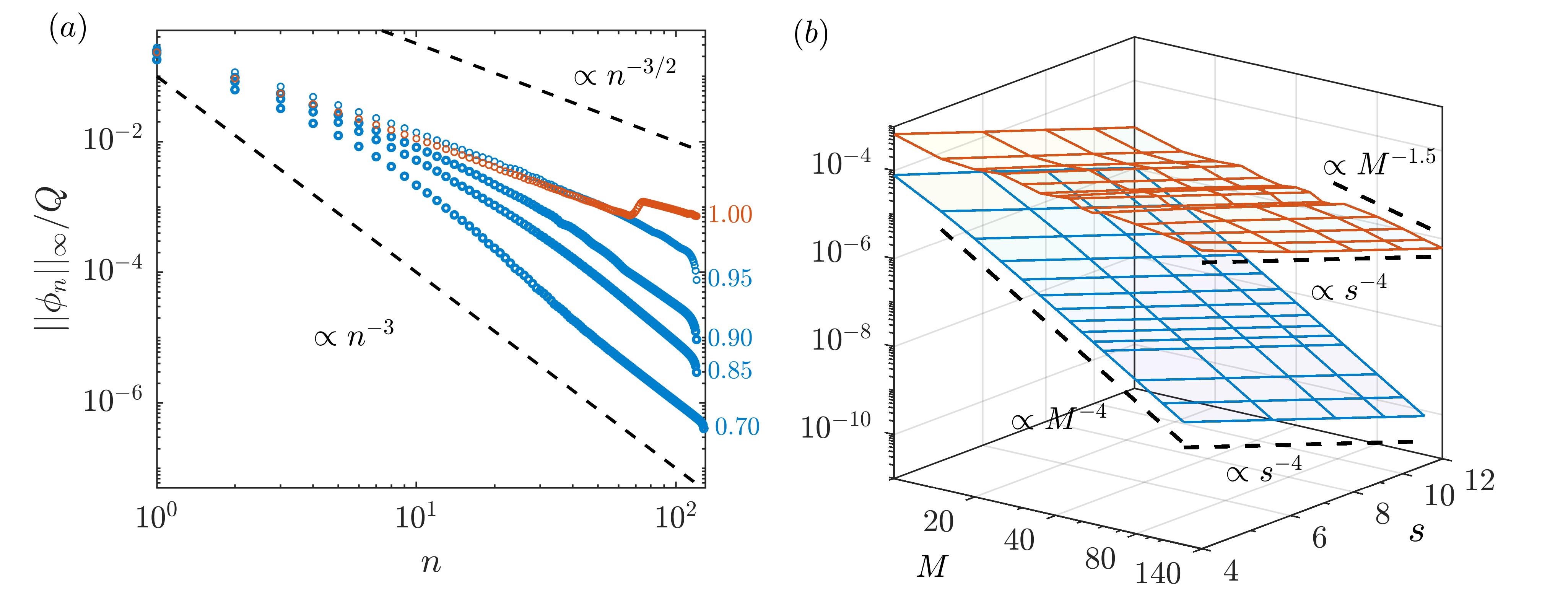} } 
\caption{$(a)$: Decay of  $\|\phi_n\|_{\infty}$ in log-scale for the case of a bump topography with $\delta = 0.5$ and $\varepsilon = \rev{0.7}$, \rev{$0.85$, $0.90$, $0.95$, and $1.00$}. $(b)$: \rev{Normalised} absolute error of the energy balance equation, \rev{$E$ = $|C_+-C_--C_{\text{int}}|/F_0$}, in log-scale as a function of the order of truncation $M$ and \rev{the} spatial discretisation parameter $s=L_M/\delta x$ for $\varepsilon=\rev{0.7}$ (blue surface) and $\varepsilon=\rev{1.0}$ (orange surface).}
\label{fig:Error_decay}
\end{figure}

We first examine the rate of decay of $\phi_n$ with $n$ obtained by solving the CMS. It is known that the baroclinic field becomes singular in the infinite-depth horizontally periodic case as $\varepsilon\rightarrow 1$ \citep{Balmforth_Ierley_Young_2002} and for a finite-depth shelf for $\varepsilon\gtrapprox 1$ \citep{griffiths2007}. In these works, the presence of the singularity is identified by a decrease in the rate of decay of the coefficients in the respective modal solutions. We verify that this is also the case for the present CMS.  We use the ``bump" profile with $\delta=0.5$, $\varepsilon\in [0.9 (0.05) 1.5]$ and solve the CMS for sufficiently large values of the parameters $(M,N_X,L_X)$ (see Section \ref{subsec:numerical_scheme}) in order to ensure that the numerical solution does not depend on them. We show the results on $\|\phi_n\|_{\infty}$ in Figure \ref{fig:Error_decay}$(b)$. We see that the decay rate drops from $n^{-3}$, the theoretically expected rate in the case of a smooth solution (Section \ref{sec:representation}), to $n^{-3/2}$ \rev{as $\varepsilon$ approaches  1}. The latter rate suggests the presence of a square root singularity on $\phi^{\dagger}$ and an inverse square root singularity on its derivatives, i.e., on the  velocities \citep{Salem1939, Raisbeck1955}.

Next, we consider the \rev{normalised} error of the energy equation \eqref{eq:energy_eq_avg}, $\rev{E} =  \lvert C_+- C_- - C_{\text{int}}\rvert/\rev{F_0}$,
where $C_{\pm}$ are calculated via \eqref{eq:Conv} and $C_{\text{int}}$ via \eqref{eq:ConvInt}. $C_{\pm}$ depends only on boundary values of $\phi_n$'s, whereas $C_{\text{int}}$ depends on their values in the entire computational domain. For an exact solution, $\rev{E}$ would be zero. Therefore, monitoring $\rev{E}$ as a function of $(M,N_X)$ is a good indication of the accuracy of the numerical solution. We use $\delta x = L_M/s$, where $L_M$ is the horizontal wavelength of the $M^{\text{th}}$ internal wave mode over the ridge
and $s= 4,6,8,10$ and $12$. We show results for $\varepsilon=\rev{0.7}$ and $\varepsilon=\rev{1.0}$
in figure \ref{fig:Error_decay}$(b)$. We obtain the \rev{expected} $s^{-4}$ decay in both cases, verifying the $4^\text{th}$-order accuracy of the spatial discretisation scheme. For $\varepsilon=\rev{0.7}$, the error decays rapidly as $M^{-4}$. \rev{As an example, we mention that for $(M,s)=(30,6)$, $E=3.1\times 10^{-7}$}. For $\varepsilon=\rev{1.0}$, the error decays as $\rev{M^{-1/2}}$ \rev{for $s\geq 6$ and $M\geq 80$}, demonstrating the slower convergence of the modal solution when the underlying field becomes singular. Nevertheless, the numerical solution accurately satisfies the energy balance even in this case. \rev{For example, for $(M,s)=(120,10)$, $E=1.6\times 10^{-6}.$} 

A final issue that must be addressed is the truncation of the infinite domain. In other words, we must ensure that the lateral boundary conditions applied at the ends of the computational domain $X$ are effective as radiation conditions and the size of $X$ does not affect the solution. If $h$ (resp., $h_x$ in the shelf case) is not compactly supported, then the length of the computational domain, $L_X$, is chosen so that $h_x$ at the boundaries is negligible and further increasing $L_X$ does not change \rev{the calculated conversion rate}. For compactly supported $h$ (resp., $h_x$), we \rev{have} examined the sensitivity of the conversion rate on the choice of $L_X$ \rev{and found that choosing $X$ as the support is sufficient for a convergent solution}. 

Concerning the choice of $M$ and $N_X$ (or $s$), some remarks are in order. As $\delta$ decreases for constant $\varepsilon$, the horizontal resolution must increase to adequately represent the topography of the ridge. On the other hand, as $\delta$ increases, the topography becomes longer and the computational domain must increase. Moreover, as $\varepsilon$ increases beyond the subcritical regime, both $M$ and the horizontal resolution must increase to achieve a good representation of the singular solution. Thus, the extent of the $(\varepsilon,\delta)$ values that can be considered depends on the given computational resources. In this work, we let \rev{$\delta$} vary in $[0.1,0.9]$, which is sufficient for our purposes. In this range, the choices $(M,s)=(64,6)$ and $(M,s)=(128,\rev{12})$ lead to convergent solutions for $\varepsilon\leq 1$ and $1\leq\varepsilon\leq 2$, respectively, \rev{while for $\varepsilon>2$ we increase $(M,s)$ further until the solution becomes independent of them}.

\section{Visualisation of flow fields}\label{sec:fields}
\subsection{Gaussian ridge}
\begin{figure}
\centering
\centerline{\includegraphics[width=\linewidth]{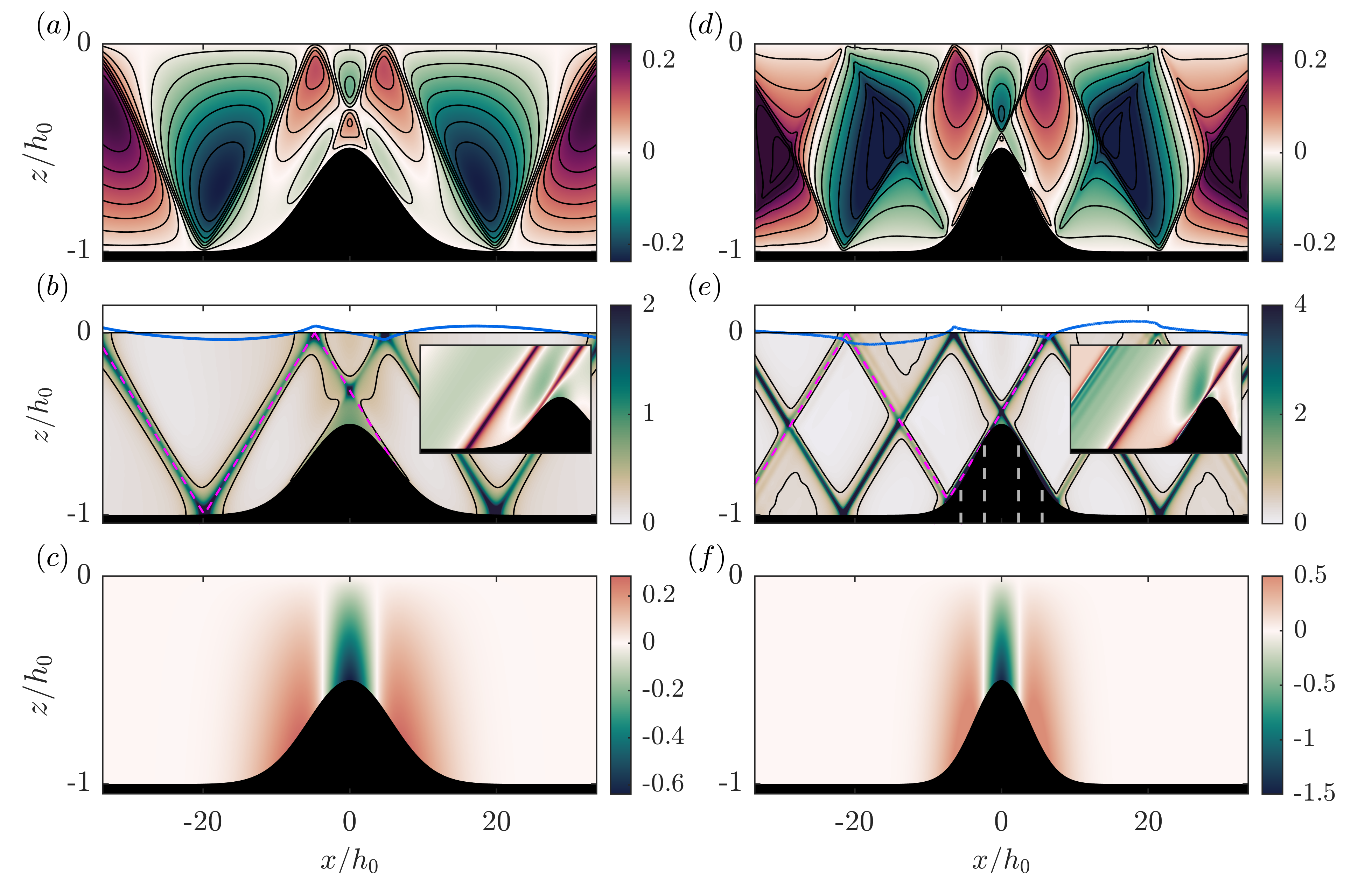}}  
\caption{Gaussian ridges with $\delta=0.5$; $\varepsilon=\rev{0.8}$ (left) and $\varepsilon=\rev{1.2}$ (right). From top to bottom, non-dimensional baroclinic stream function $\psi^{\#}/Q$, energy density $\mathcal{E}^{\#}/U_0^2$, and \rev{body} forcing term $\varepsilon^2b^2\Phi_{xx}^{(0)}/Q$. Free-surface elevation $3\times10^4\eta/h_0$ is also shown in light blue. \rev{Black lines correspond to iso-energy curves of $\mathcal{E}^{\#}/U_0^2 = 0.36$.} \rev{Magenda} dashed lines correspond to characteristic lines of slope $\pm \mu^{-1}$. Vertical dashed lines correspond to the critical points of the topography. Insets show \rev{the field $\phi^{\#}_x+\phi^{\#}_z/\mu$  in  the region $[-30,+7] \times [-1.05,0]$.} \rev{The colormaps used in this and the subsequent figures are from \cite{Thyng}.} }
\label{Fig:flowmaps_Gaussian}
\end{figure}

Here we consider solutions of the CMS for a subcritical ($\varepsilon=0.\rev{8}$) and a supercritical ($\varepsilon=1.\rev{2}$) Gaussian ridge. In figure \ref{Fig:flowmaps_Gaussian}, we show the purely baroclinic stream function $\phi^{\#}$, the 
energy density $\mathcal{E}^{\#}= (\mathbf{u}^{\#})^2/2 +(b^{\#})^2/2/N^2$, the reconstructed free-surface elevation, and the body-forcing term $\Phi^{(0)}_{xx}$.

We clearly observe the beam\rev{-like} structure of ITs, which is finer and more intense in the supercritical case. In this case, the solution approximates a field that is continuous along the beams, where the energy density attains very large values (theoretically infinite). The free surface also transforms from smooth to cusp-like as $\varepsilon$ exceeds $1$. Such fine-scale large-amplitude features are regularised in the ocean via 
non-linearity and viscosity, which are beyond the present theory. Nevertheless, such flow-field calculations are readily obtained using the CMS and give us useful information for observable quantities. For example, in the supercritical case, the global maximum on the horizontal velocity is attained at the point of intersection of the beams above the ridge. The following maxima are attained at the first points of reflection at the free surface. Right at these points, the free surface attains its maximum slope.

Interestingly, \rev{the solution varies considerably between the sloping regions and the flatter regions.} 
This is due to the body-forcing term in \eqref{eq:BVP_freq0H}, which couples the baroclinic motion with the non-uniform barotropic current and scales with $\rev{(1/h)_{xx}}$ (Figures \ref{Fig:flowmaps_Gaussian}$(c)$ and $(f)$). This effect becomes more apparent if we write the PDE in \eqref{eq:BVP_freq0H} as a first-order system,
\begin{align}\label{eq:first_order}
\left(\phi^{\dagger}_x\mp\frac{1}{\mu}\phi^{\dagger}_z\right)_x\pm\frac{1}{\mu}\left(\phi^{\dagger}_x\mp\frac{1}{\mu}\phi^{\dagger}_z\right)_z &= -\Phi^{(0)}_{xx}.
\end{align}
\rev{Note that the above equation also holds with $\phi^{\dagger}$ replaced by $\phi^{\#}$ and $\Phi^{(0)}$ by $\Phi$.} Eq.\ \eqref{eq:first_order} shows that the solution is not constant along a given characteristic  unless $\Phi^{(0)}_{xx}=0$. \rev{This is captured by our solution as an adjustment of the iso-energy curves near the ridge crest, figures \ref{Fig:flowmaps_Gaussian}$(b)$ and $(e)$. In the insets of these figures, we also plot the field $\phi^{\#}_x+ \phi^{\#}_z/\mu$ in order to highlight the variation of the beams.}\rev{\ref{Fig:flowmaps_Gaussian})}. This effect seems to have gone unnoticed in the literature, although it is clearly present in the calculations of \cite{Stashchuk}. It can also be inferred from the calculations of \cite{Baines1973}, even though he does not visualise flow fields in the physical domain.

\subsection{A non-radiating ridge}
\begin{figure}
\centering
\centerline{\includegraphics{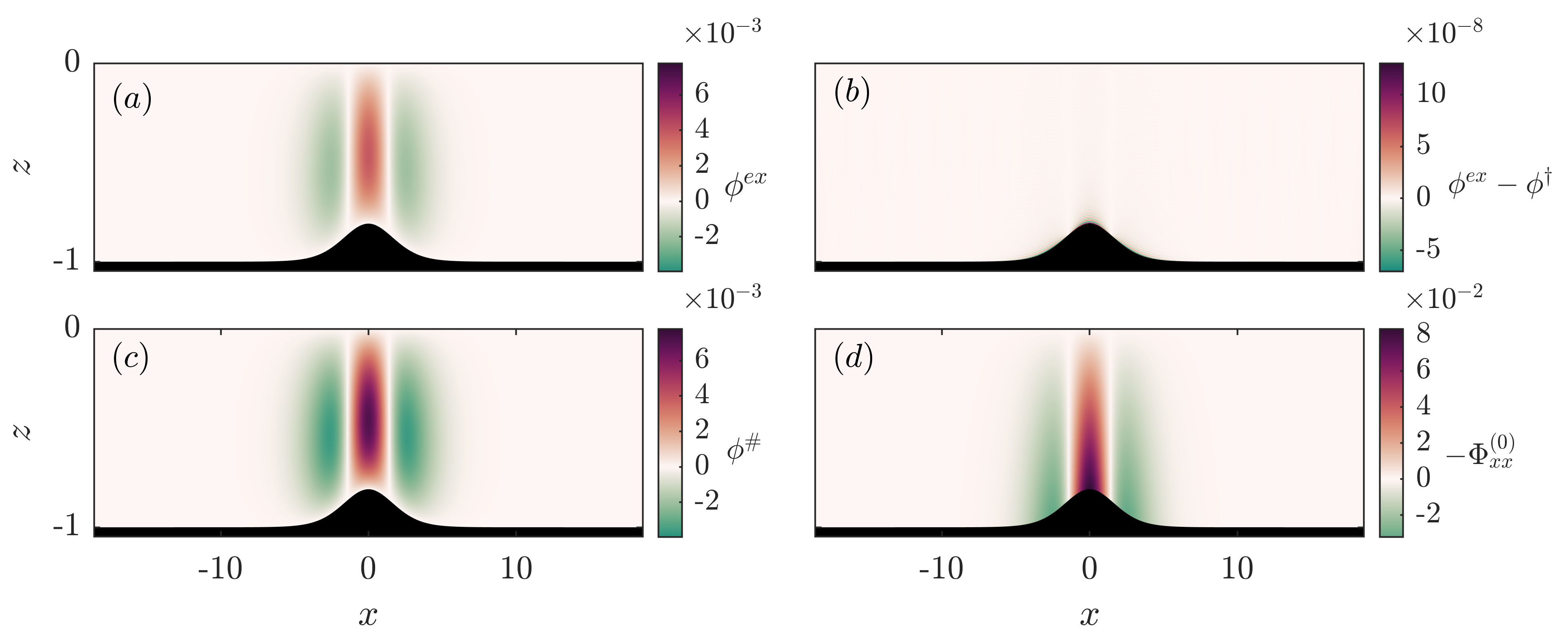}} 
\caption{The non-radiating ridge of \cite{maas_2011}; Non-dimensional $(a)$ exact baroclinic response $\phi^{ex}$, and $(b)$ \rev{difference $\phi^{ex}-\phi^{\dagger}$ where $\phi^{\dagger}$ is calculated using the CMS}, $(c)$ purely baroclinic response $\phi^{\#}$ calculated with the CMS, \rev{and} $(d)$ body-forcing term $\Phi^{(0)}_{xx}$. All variables are dimensionless.}
\label{Fig:Maas}
\end{figure}

For further validation, we reproduce a case investigated by \citet[Section 4.2, ``Ridge'']{maas_2011}. Maas considered the dimensionless version of Eq.~\eqref{eq:BVP_freq0H}, obtained by the scaling $(x,z,\phi^{\dagger},h) = (L\tilde{x},L/\mu\tilde{z},Q\tilde{\phi^{\dagger}},L/\mu\tilde{h})$, and showed that it has an exact solution, denoted here by $\phi^{ex}$, that radiates no energy at infinity for a specially constructed ridge; see also \cite{wunsch_wunsch_2022} for an alternative construction of non-radiating topographies. For the case 
in Eq.\! (4.15) of \cite{maas_2011}, which is very similar to a Gaussian ridge with $\Lambda=0.19$ and $L=e/\sqrt{2}$ in our Eq.(\ref{eq:topos}), the exact non-radiating solution $\phi^{ex}$ is shown in Figure \ref{Fig:Maas}$(a)$. \rev{Our calculation, $\phi^{\dagger}$, is in agreement with $\phi^{ex}$ as shown in figure \ref{Fig:Maas}$(b)$ where the difference $\phi^{ex}-\phi^{\dagger}$ is plotted.} The field $\phi^{ex}$ (or $\phi^{\dagger}$) is a  combination of a baroclinic and a non-hydrostatic barotropic response trapped near the ridge. \rev{The purely baroclinic response is shown in Figure \ref{Fig:Maas}$(c)$}. The hydrostatic body-forcing term is shown in Figure \ref{Fig:Maas}$(d)$. 
The calculated non-dimensional energy conversion rate is on the order \rev{$10^{-17}$}. 

\section{Energy conversion rates}\label{sec:energy}
In this section, we calculate the total energy conversion rate $C=C_+-C_-$ (Eq.~\ref{eq:energy_eq_avg}) for the topographies presented in Section \ref{sec:Validation}. Our primary objective is to examine the dependence of $C$ on $\varepsilon$. We also examine the region of validity of the WTA prediction (Eq.~\ref{eq:ConversionStL}).

\begin{figure}
\centerline{
\includegraphics[width= 14cm]{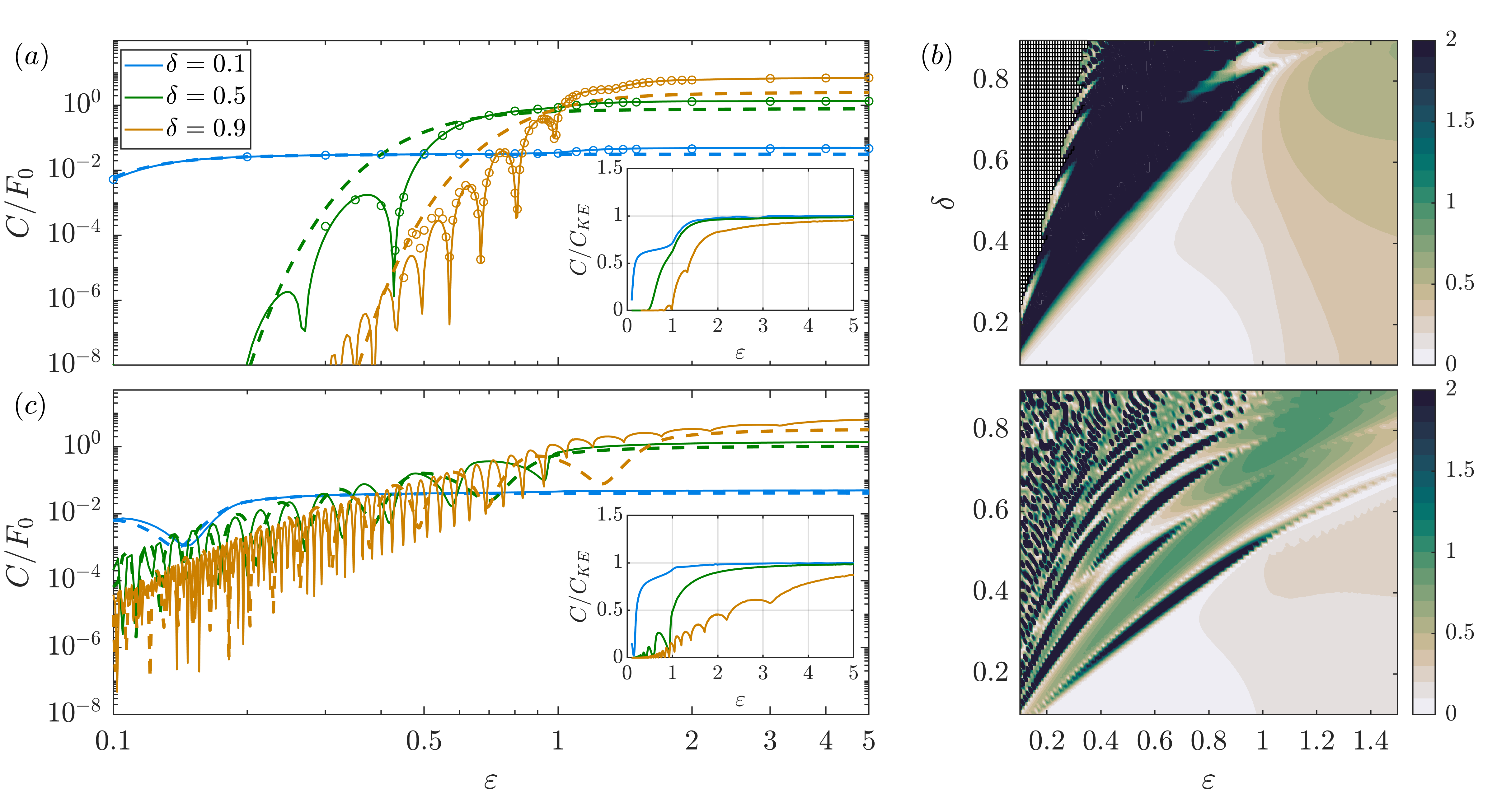}}
\caption{$(a)$ Calculated non-dimensional energy conversion rate $C/F_0$ in log-log scale as a function of $\varepsilon$ (solid lines) for $\delta=0.1$, $0.5$ and $0.9$, for a Gaussian ridge. Dashed lines correspond to the WTA prediction $C^{\text{WTA}}/F_0$. \rev{Circles correspond to the results obtained by using the Green's function method.} The inset corresponds to the same calculation in linear scale, normalised by $C_{KE}$ which denotes the knife-edge prediction of \cite{llewellyn_smith_young_2003}. $(b)$ Relative error $|(C-C^{\text{WTA}})/C|$ for $(\varepsilon,\delta)\in[0.1,1.5]\times[0.1,0.9]$. The same calculations for the bump ridge are shown in $(c)$ and $(d)$. The hatched region in $(b)$ corresponds to topographies for which conversion is negligible for both the full solution and the WTA.}
\label{Fig:Conversionridges}
\end{figure}

We first consider Gaussian profiles with $\delta=0.1$, $0.5$, $0.9$, and $\varepsilon$ ranging from $0.1$ to $5$. We show in Figure \ref{Fig:Conversionridges}$(a)$ the non-dimensional calculated conversion rate $C/F_0$, where $F_0$ is given in \eqref{eq:ConversionStL}, using a log-log plot in order to capture large relative variations. In the inset, we show $C/C_{KE}$, where $C_{KE}$ is the knife-edge prediction of \cite[Eq. (2.12)]{llewellyn_smith_young_2003}, using a linear plot to focus on the largest $\varepsilon$ values.  For $\delta=0.1$, $C$ increases rapidly up to about $\varepsilon=0.2$ and more slowly afterwards. The WTA prediction is in excellent agreement with the full solution in the entire subcritical regime. For $\delta=0.5$ and $0.9$, we clearly observe abrupt drops in $C$ for discrete values of $\varepsilon$.
However, note that $C/F_0$ never strictly drops to zero. Nevertheless, calculations for values of $\varepsilon$ closely straddling a local conversion minimum in the case $\delta=0.5$ reached conversion values approaching machine precision (not shown). The WTA presents strong qualitative and quantitative differences and clearly fails to predict these abrupt drops. As $\varepsilon$ exceeds 1, the transition looks smooth \rev{for all $\delta$}. \rev{For $\delta=0.1$, $C$ is initially increasing until $\varepsilon\approx 2.0$, after which it remains very close to $C_{KE}$}. \cite{Khatiwala2003} also reports this increase for a small-height Gaussian profile up to $\varepsilon=1.6$ in his non-linear calculations. \rev{For $\delta=0.5$, $C$ monotonically increases with $\varepsilon$, approaching $C_{KE}$.} \rev{For $\delta=0.9$, a local minimum is attained at about $\varepsilon = 1.4$, after which $C$ increases, also approaching $C_{KE}$} (inset of figure \ref{Fig:Conversionridges}$(a)$).  It should also be noted that we obtained a similar trend for the Witch of Agnesi profile (not shown). \rev{In figure \ref{Fig:Conversionridges}$(a)$, we also show that the calculated conversion rate is in overall agreement with the one obtained by using the Green's function method \citep{echeverri_peacock_2010,iTides}. The differences appearing for small $\varepsilon$ in the case $\delta=0.9$ are due to the domain truncation needed to avoid singularities in the Green-function solution. Furthermore, the modal convergence being slower than the CMS approach, these values are also less precise\footnote{\rev{The horizontal domain considered describes between $99\%$ and $99.5\%$ of the variation of height of the topography.}}.
We refer to section \ref{sec:discussion} for a further discussion.}   

We also performed the same calculations as above for the bump profile, figure \ref{Fig:Conversionridges}$(b)$. For $\delta=0.1$, $C/F_0$ reaches a single local minimum, which is adequately predicted by the WTA. For larger $\delta$, similarly as before, we observe local extrema in the full solution, although in this case there are many more in any given range of $\varepsilon$ values. The WTA gives qualitatively similar results, predicting certain local extrema with values comparable to the full solution, but not at the right locations. A similar behaviour is also reported in \cite[Section 2.3.1]{Vlasenko} in terms of the amplitude of the first mode for a small-height sinusoidal bump. As in the Gaussian profile, the transition to the supercritical regime is smooth and \rev{for larger $\varepsilon$ and all $\delta$ the conversion rates approach $C_{KE}$.} \rev{In the extreme case $\delta=0.9$, local extrema in $C$ are obtained up to $\varepsilon = 3.2$. After this value, $C$ increases slowly towards $C_{KE}$. In this case, we have extended our calculations up to $\varepsilon=11$ and found that $C/C_{KE}\approx 0.98$.}

In order to assess the region of validity of the WTA prediction, we performed more than 20,000 calculations per ridge in the parameter space $(\varepsilon,\delta)=[0.1,1.5]\times[0.1,0.9]$ and calculated the relative error $|(C-C^{\text{WTA}})/C|$. We show the results in the right panels of Figure~\ref{Fig:Conversionridges}. Note that in the case of the Gaussian ridge, there is a hatched region (Figure~\ref{Fig:Conversionridges}$(b)$, upper left corner) that corresponds to a practically negligible conversion rate $C/F_0$ on the order of $10^{-8}$. The corresponding values obtained by the WTA reach machine precision. Based on these figures, it is clear that in both cases we may identify coherent regions in the parameter space for which the WTA is valid according to some relevant criterion (e.g., conversion rates that differ by less than 10\% with respect to the full solution). Interestingly, these regions include unexpectedly large values of $\varepsilon$ and $\delta$. However, it should be stressed out that the region of validity of the WTA depends strongly on the type of the ridge and is clearly smaller in the case of the bump profile.    

\begin{figure}
\centerline{\includegraphics[scale = 0.8]{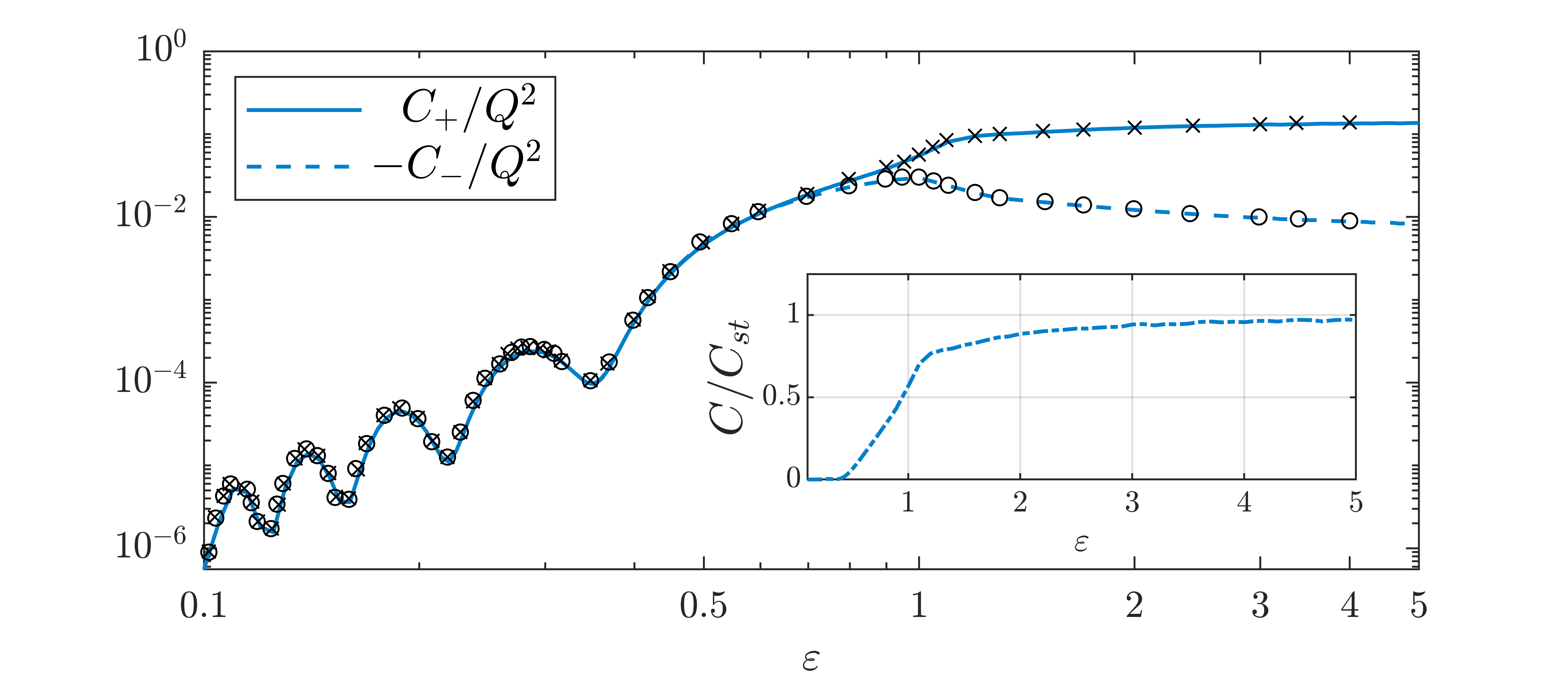}}
\caption{Non-dimensional \rev{conversion rates} $ C_-/Q^2$ (\rev{blue solid line}) and $C_-/Q^2$ (\rev{blue dashed line})  for the shelf profile with $h_-/h_+ = 0.5$. Circles and crosses correspond to the calculations of \cite{griffiths2007}. \rev{The inset
corresponds to the same calculation in linear scale and the dotted dashed blue line shows the total calculated conversion rate, $C=C_+-C_-$,  normalised by \cite{StLaurent2003}'s prediction  for a step profile, $C_{st}$.}}
\label{Fig:CGrif}
\end{figure}

We close this section by demonstrating the applicability of the CMS in the case of a shelf profile. We consider the topography in Eq.~\eqref{eq:shelf_topo} with $h_-=2000$~m and $h_+=1000$~m (\rev{$\delta = 0.5$}) for \rev{which} calculations based on another modal decomposition \rev{method} are available in \citet{griffiths2007}, GG07, for $\varepsilon\in[0.1,4]$. As in that work, we use the hydrostatic version of the CMS (Remark \ref{rem:HA}).  We plot our results on the conversion rates $C_{\pm}$ \rev{for $\varepsilon\in[0.1,5]$} in figure \ref{Fig:CGrif}, together with the results digitised from Fig.~11 of GG07. \rev{The results from the two methods are in  agreement for all $\varepsilon$.
In the inset of figure \ref{Fig:CGrif}, we additionally compare in a linear scale our calculations with the conversion rate for a step profile, $C_{st}$, obtained  by using eigenfunction matching at the topographic discontinuity \citep{StLaurent2003}\footnote{\rev{We obtained this result from Eqs.\ (32)-(36) of} \cite{StLaurent2003} \rev{for 2000 modes. Note however that for $\delta=0.5$, the matrix in Eq.\ (32) is singular and a value $\delta=0.500001$ is used here.}}. As is also noted in GG07, $C$ is slowly increasing up to $\varepsilon=4$. This increase is also reported in the early calculations of \cite{Craig1987} for a linear step up to $\varepsilon=2$. Our calculations for $\varepsilon>4$  suggest that $C$ slowly approaches the value $C_{st}$ from below. In fact, we have observed that $C/C_{st}$ goes from 0.98 for $\varepsilon=5$, to 0.99 for $\varepsilon=10$. }

\section{Discussion on the Green's function method\label{sec:discussion}}
We briefly discuss here how our method compares with an existing method for IT calculations based on \cite{Robinson1969}'s Green's function 
for the homogeneous internal wave equation in a uniform strip \citep{Petrelis_et_al_2006, echeverri_peacock_2010}. This approach is limited to domains of the same depth at infinity, say $h_0$, and the starting point is to write the total flow as $\phi^{\text{G}}(x,z) =\Phi_0(z) + \vartheta(x,z)$,
where $\Phi_0=-Qz/h_0$ represents the barotropic flow corresponding to the uniform strip $[-\infty,+\infty]\times[-h_0,0]$. 
From \eqref{eq:BVP1_freq}, the field $\vartheta$ solves
 \begin{align}\label{eq:BVP_Green}
\Lin_{\mu}\vartheta  = 0,\quad
\vartheta(x,0)=0\quad\text{and}\quad
\vartheta(x,-h)= Q(1-h/h_0),  
\end{align}
together with conditions at $x\rightarrow\pm\infty$.
Next, in order to solve \eqref{eq:BVP_Green}, $\vartheta$ is expressed as an ansatz defined as the integral of the product of the Green's function evaluated at $z=-h$ with an unknown distribution $\gamma(x)$,
\begin{align}\label{eq:Green_theta}
    \vartheta(x,z)=\int_{-a}^{a}\gamma(x')G(x,x',z,-h(x'))\dd{x'},
\end{align}
where it is additionally assumed that $-h_0+h$ is supported on $[-a,a]$. Note that \eqref{eq:Green_theta} implies that $\vartheta$ is defined on $[-a,a]\times[-h_0,0]$  therefore trench topographies (for which $h\geq h_0$) are also excluded. 
Expression \eqref{eq:Green_theta}  satisfies the first two equations in \eqref{eq:BVP_Green} as well as the radiation conditions as $a\rightarrow+\infty$. The third equation in \eqref{eq:BVP_Green} yields
\begin{align}\label{eq:Fredholm}
     \int_a^{-a}\gamma(x')G(x,x',-h(x),-h(x'))\dd{x'} =  Q(1-h/h_0),
\end{align}
which must be solved for $\gamma(x)$. Eq.\! \eqref{eq:Fredholm} is a Fredholm integral equation of the first kind, which is known to be ill-posed (see e.g.\ \citet[Section 1.1]{Groetsch}, \cite{Groetsch2}).
This has unpleasant consequences, as is recognised by \cite{Petrelis_et_al_2006} and \cite{echeverri_peacock_2010}; the linear system obtained via the series representation of the discontinuous Green's function and the spatial discretisation of Eq.~\eqref{eq:Fredholm} becomes ill-conditioned or even singular if $a$ is large or if $h_x\approx 0$ in some region, among other special cases \citep{echeverri_yokossi_balmforth_peacock_2011}. Moreover, as noted by \cite{maas_2011}, 
the physical meaning of $\vartheta$ is clear only as $\rev{x\rightarrow\pm\infty}$, in which case $\vartheta(x,-h)\rightarrow 0$ and $\vartheta$ may be interpreted as the far-field baroclinic response.
Therefore,  the present CMS is more general and numerically advantageous. There exist, however, two issues that should be discussed further.

The first issue concerns the \rev{ local dynamics} above the topography. As shown in Section \ref{sec:fields}, the \rev{baroclinic wave field varies significantly in regions where} the body-forcing term \rev{is important (recall Eq.~\eqref{eq:first_order})}. On the other hand, the ansatz \eqref{eq:Green_theta} postulates that $\vartheta$ can be represented as a superposition of sources, all of which ignore the interaction between the baroclinic and the non-uniform barotropic background flow. The intensities of the sources are instead determined by an ill-posed equation stemming from the bottom boundary condition in \eqref{eq:BVP_Green}. \rev{We can reconcile the two approaches by verifying that the purely baroclinic field, satisfying Eq.\ \eqref{eq:BVP_freq0NH}, is recovered by $\phi^{\#} = \Phi_0+\vartheta-\Phi$, with $\Phi = \Phi^{(0)}+\Phi^r$.}

The second issue is associated with the  the so-called ``knife-edge limit'' \rev{established in \cite{Petrelis_et_al_2006}, see also \cite{echeverri_peacock_2010}}. The conversion rate obtained via the ansatz \eqref{eq:Green_theta} tends to this limit because
\eqref{eq:Green_theta} is actually a direct extension of the knife-edge solution of \cite{llewellyn_smith_young_2003} in the smooth topography case. In fact, \cite{Petrelis_et_al_2006} state that only in the knife-edge case is ``the kernel of the integral equation (\dots) so singular that the resulting linear system is well conditioned''.

The knife-edge solution corresponds to a discontinuous topography ($\varepsilon=\infty$) for which the body-forcing term in Eq.~\eqref{eq:BVP_freq0H} (or, equivalently, in Eq.~\eqref{eq:BVP_freq0NH}) is not well-defined because Eq.\! \eqref{eq:BarotropicBVP} requires the domain to be smooth. Moreover, in the knife-edge formulation, the amplitude of the horizontal barotropic velocity is constant, say $U_0$. This implies that the total flux through a vertical cross-section is $U_0h_0$ everywhere except \rev{at} the position of the knife-edge where it equals $U_0(h_0-\Lambda)$, with $\Lambda$ the height of the knife-edge. This is not consistent with the key assumption in the body-forcing formulation that the barotropic flow accounts for a constant total flux through any vertical cross-section. Therefore, \rev{the present numerical evidence showing that the conversion rate, obtained by the body forcing formulation, approaches the knife-edge (or step) limit for large $\varepsilon$ seems unexpected. Although this limiting behaviour of the singular solution is interesting, its theoretical analysis is beyond the scope of this paper and remains a topic for future investigation.}

\section{Conclusions\label{sec:conclusions}}
We developed a new infinite CMS describing the generation of linear ITs in 2D using \cite{Garett_Gerkemma_2007}'s body-forcing formulation and an exact, local eigenfunction expansion of the stream function. In the weak topography limit, we recover from this CMS the classical formula of \cite{Llewellyn2002} for the conversion rate. For general topographies, we solve the truncated CMS numerically using fourth-order finite differences and demonstrate the convergence and accuracy of the semi-analytical scheme. The formation of singularities is detected as a decrease in the rate of decay of the modal amplitudes.
We additionally derive the energy equation for the body-forcing formulation and show that our solution verifies it with very good accuracy, even when the field becomes singular.  Further, we show how one can obtain the purely baroclinic response and also propose a method to estimate the free-surface elevation induced by the baroclinic motion within the rigid-lid approximation.

 By reconstructing the flow field, we showed that the interaction between ITs and the background non-uniform barotropic flow \rev{affects the dynamics} locally around the sloping topography. We calculated the conversion rate for two different ridges. For the commonly used Gaussian profile, we find distinct points of practically zero energy radiation in the subcritical regime for $\delta>0.1$. We have checked that similar null-points exist also for the Witch of Agnesi profile (not shown). Our calculations verify the hypothesis of \cite{maas_2011} that non-radiating topographies are common and possible for subcritical topographies with sufficiently large values of $\delta$. 
  For the compactly supported bump profile, local extrema of the conversion rate exist for all $\delta$ and persist even in the supercritical regime \rev{for large $\delta$}. These local minima  do not correspond to null points, but rather to points of weak radiation. \rev{These differences are due to the qualitative differences between a Gaussian and a bump profile. The baroclinic response depends not only on the profile, but also on its first and second derivatives through the body forcing term Eq.\ \eqref{eq:BVP_freq0H}, which can differ significantly for an infinite-support decaying topography and a compactly supported one. In the context of the WTA approximation, the different behaviour of the conversion rates can also be explained by the different decays of the Fourier transforms of the topographies. We have also tested the Witch of Agnesi profile and a compactly supported polynomial ridge and found that these two different behaviours persist. Thus, it appears that the trend of the conversion rate for a Gaussian ridge (resp. bump ridge) is typical for infinite-support (resp. compactly supported) ridges.} 
We have also quantified the error of the WTA with respect to the full linear solution. Our general conclusion is that the WTA is adequate even for large $\varepsilon$ for sufficiently small $\delta$ but the region of its validity strongly depends on the type of topography.
\rev{Similarly, our results show that the knife-edge limit agrees with the linear singular solution for sufficiently large $\varepsilon$, although the value of $\varepsilon$ for which this limit is attained depends on $\delta$ and the topography.} 

\rev{Our comparisons show that the CMS and the Green's function methods yield the same conversion rate for ridges. However, the CMS can be applied to more general domains and is numerically more convenient. The solutions fields of two methods differ over sloping regions, the solution from the Green's function method has to be corrected to properly describe the purely baroclinic field. This may be of importance for the horizontal phase propagation of internal tides above topographies, and could have an influence for sea-surface observations and analysis.}\rev{Our calculations in the case of a shelf agree }  with the modal solution of \cite{griffiths2007}. \rev{The main advantage of the present CMS is that it exhibits faster convergence, thus it is numerically more efficient. By extending our calculations to the strongly supercritical regime we provide new numerical evidence that the conversion rate approaches the one obtained in the case of a step \citep{StLaurent2003}}.  
 
The present CMS can be extended in several directions. \rev{It can be applied, for instance, to the investigation of topographically trapped flows for arbitrary topographies extending the study of \cite{Maas1989}}. The assumption of constant stratification can be easily lifted, at the cost of using a set of local basis functions that is calculated numerically \rev{in the case of smooth stratification}.  The linearised free-surface condition can be taken into account by using basis functions defined in terms of a local transcendental equation.
The solution of the CMS for other topographies such as trenches, multiple ridges or realistic topographic profiles is straightforward. \rev{A Matlab script implementing the solution of the CMS is freely available through the following link \href{https://github.com/chpapoutsellis/InternalTidesCMS}{https://github.com/chpapoutsellis/InternalTidesCMS} .}

\section*{Declaration of Interests}
The authors report no conflict of interest.
\section*{Acknowledgments}
The authors acknowledge the support of the IRT/STAE Foundation (funding for the post-doctoral contract of Ch.P.), and also L.R.M.\ Maas for fruitful discussions. \rev{We would like to thank the reviewers and J. Nycander for their comments and suggestions, which greatly improved the quality of this paper.} M.M.\ thanks Lucile Planes and Martin Benebig for testing computations with iTides. N.G.\ acknowledges the support of the Natural Sciences and Engineering Research Council of Canada (NSERC) [funding reference numbers RGPIN-2015-03684 and RGPIN-2022-04560] and of the Canadian Space Agency [14SUSWOTTO].

\appendix
\section{Asymptotic analysis of the barotropic equations}\label{app:barotropic}

The barotropic flow solves the system \citep{Baines1973, Garett_Gerkemma_2007}
\begin{subequations}\label{eq:Euler_Barotrop}
	\begin{align}
	U_t - f V			& =  -P_x,\quad V_t + f U=  0, \label{eq:Eulx_Barotrop} \\
	W_t				    & =  -P_z, \label{eq:Eulz_Barotrop}\\
	B_t + N^2 W		& =  0, \label{eq:Buoy}\\
	U_x  + W_z	& =  0,\label{eq:incompr_Barotrop}
    \end{align}
\end{subequations}
where $(U,V,W)$, $P$ and $B$ denote respectively the velocities, the scaled pressure, and the buoyancy induced by the barotropic flow. On the boundaries, we have
\refstepcounter{equation}
\[
h_x U(x,-h)+W(x,-h)  = 0,\quad W(x,0)  = 0.
\eqno{(\theequation{\text{a},\text{b}})}\label{eq:BCs_Barotrop}
\]
As described in section \ref{subsec:barotropic}, the above system reduces to the BVP \eqref{eq:BarotropicBVP} in terms of $\Phi$. In order to derive an asymptotic solution for \eqref{eq:BarotropicBVP} we introduce the scaling,
\begin{align}\label{eq:scaling}
\tilde{z}=z/h_0,\quad \tilde{x} = x/L,\quad \tilde{h}=h/h_0,\quad \tilde{\Phi} = \Phi/Q,
\end{align}
and  the parameter $\sigma = h_0^2/L^2$,
where $L$ is the horizontal scale of the topography.  The dimensionless version of \eqref{eq:BarotropicBVP} is written as, after dropping the tildes,
\begin{align}\label{eq:BarotropicPDE_ND}
\sigma\Phi_{xx} + \mu_0^{-2}\Phi_{zz}=0,\quad
\Phi(x,0)=0,\quad
\Phi\left(x,-h\right)= 1.
\end{align}
If $\sigma\ll1$, we can solve \eqref{eq:BarotropicPDE_ND} by using the asymptotic expansion $\Phi^{\text{app}} = \sum_{i=0}^{K}\sigma^i\Phi^{(i)}$,
for some $K\geq 0$. Substitution of $\Phi^{\text{app}}$ in the first equation of \eqref{eq:BarotropicPDE_ND} yields
\begin{align}\label{eq:Phi_i_zz}
\sum_{j=0}^{K}\sigma^{j}\left(\Phi^{(j-1)}_{xx}+\mu_0^{-2}\Phi^{(j)}_{zz}\right)&=O(\sigma^{K+1}).
\end{align}
where the convention $\Phi^{(-1)}=0$ is used. By requiring the residual $O(\sigma^{K+1})$ to be cancelled we obtain
the following recurrence relation:
\begin{align}\label{eq:rec}
\begin{aligned}
(j=0):&\quad \Phi^{(0)}_{zz} =0,\quad \Phi^{(0)}(x,0)=0,\quad \Phi^{(0)}(x,-h)=1\\
(1\leq j\leq K):&\quad\begin{aligned}\Phi^{(j)}_{zz} &= -\mu_0^2\Phi^{(j-1)}_{xx},\quad 
\Phi^{(j)}(x,0)&=0,\quad 
\Phi^{(j)}(x,- h)&=0.\end{aligned}
\end{aligned}
\end{align}
Solving the above BVPs we obtain 
\begin{align}\label{eq:Phi_orders_nondim}
\begin{aligned}
\Phi^{(0)} = \frac{z}{- h},\,\,\,\,\Phi^{(j)} = \frac{z\mu_0^2}{ h}\int_{-h}^{0}\int_{- h}^{z'}\Phi^{(j-1)}_{xx}\dd{z''}\dd{z'} + \mu_0^2\int^{0}_{z}\int_{- h}^{z'}\Phi^{(j-1)}_{xx}\dd{z''}\dd{z'}.
\end{aligned}
\end{align}
Performing the computation for $j=1$ and using \eqref{eq:scaling}  we find
\begin{align}\label{eq:Phi_orders_dim}
\Phi^{(0)}=Q\frac{z}{-h},\quad
\Phi^{(1)}=-Q\mu_0^2\left(\frac{1}{-h}\right)_{xx}\frac{1}{6}\left(z^3-h^2z\right).
\end{align}

We turn now to the system
\eqref{eq:Euler_Barotrop}--\eqref{eq:BCs_Barotrop}. Introducing the scaling $\tilde{t} = \omega t$ and
\begin{align}
    \tilde{U} =U/U_0,\quad  \tilde{V} = V/U_0,\quad \tilde{W} = W/W_0, \quad \tilde{P} = P/(\omega U_0 L), \quad \tilde{B}= B/g,
\end{align}
with $U_0=\sqrt{g h_0}$ and $W_0 = U_0h_0/L$, \eqref{eq:Euler_Barotrop}--\eqref{eq:BCs_Barotrop} become, after dropping the tildes,
\begin{subequations}\label{eq:Euler_Barotrop_ND}
	\begin{align}
	U_t - \frac{f}{\omega} V			& =  -P_x,\quad V_t + \frac{f}{\omega} U=  0, \label{eq:Eulx_Barotrop_ND} \\
	\sigma W_t				    & =  -P_z, \label{eq:Eulz_Barotrop_ND}
    \end{align}
\end{subequations}
\begingroup\abovedisplayskip=0pt
\begin{align}	B_t + \frac{N^2}{\omega g} W		& =  0,\quad\quad\quad\,\,\, \label{eq:Buoy_ND}
\end{align}
\begin{align}
	U_x  + W_z	& =  0,\quad\quad\,\, \label{eq:incompr_Barotrop_ND}
\end{align}
\begin{align}\label{eq:BCs_Barotrop_ND} 
 h_x U(x,- h)+W(x,- h)  = 0,\quad
 W(x,0)  = 0. 
\end{align}
\endgroup
Plugging $\Xi^{\text{app}}=\sum_{i=0}^{K}\sigma^{i}\Xi^{(i)}$, where $\Xi$ is any of the fields $U$, $V$, $W$, $B$, $P$,
into \eqref{eq:Euler_Barotrop_ND}--\eqref{eq:BCs_Barotrop_ND}, we see that $\Xi^{(i)}$
satisfies  \eqref{eq:Eulx_Barotrop_ND}, \eqref{eq:Buoy_ND}--\eqref{eq:BCs_Barotrop_ND}
while \eqref{eq:Eulz_Barotrop_ND} gives
\begin{align*}
    \sum\limits_{i=0}^{K}\sigma^{i}\left(W^{(i-1)}_t+P_z^{(i)}\right) = O(\sigma^{K+1}).
\end{align*}
Then $O(\sigma^{K+1})$ is cancelled if $W^{(i-1)}_t+P_z^{(i)} = 0$,
and in dimensional form we have
\begin{subequations}\label{eq:Euler_Barotrop_asympt}
	\begin{align}
	U^{(i)}_t - f V^{(i)}			& =  -P^{(i)}_x,\quad V^{(i)}_t + f U^{(i)}=  0, \label{eq:Eulx_Barotrop_asympt} \\
	W^{(i-1)}_t				    & =  -P^{(i)}_z, \label{eq:Eulz_Barotrop_asympt}
    \end{align}
\end{subequations}
\begingroup\abovedisplayskip=0pt
\begin{align}	B^{(i)}_t + N^2 W^{(i)}		& =  0,\quad\quad\quad\,\,\, \label{eq:Buoy_asympt}
\end{align}
\begin{align}
	U^{(i)}_x  + W^{(i)}_z	& =  0,\quad\quad\,\, \label{eq:incompr_Barotrop_asympt}
\end{align}
\begin{align}\label{eq:BCs_Barotrop_asympt}
h_x U^{(i)}(x,-h)+W^{(i)}(x,-h) = 0,\quad
 W^{(i)}(x,0)  = 0. 
\end{align}
\endgroup
For $i=0$, Eq.\ \eqref{eq:Eulz_Barotrop_asympt} is $ P^{(0)}_z = 0$, thus, at leading order the flow is hydrostatic.

\section{Matrix coefficients of the CMS}\label{app:ABC}
The matrix coefficients appearing in \eqref{eq:CMS} are given by
\begin{align}\label{ABCcoeffs_explicit}
	b_{m n}   &=  \left\{\begin{aligned}4(-1)^{m+n}\frac{m n}{m^2-n^2},\, m\neq n\\ 1,\, m= n\end{aligned}\right.,\\
	c_{m n}  & = \left\{\begin{aligned}-4(-1)^{m+n}\frac{mn(m^2+n^2)}{(m^2-n^2)^2},\,  m\neq n\\
	-\frac{1}{2}-\frac{1}{3}n^2\pi^2,\,  m= n \end{aligned}\right.\\
  d_{m n}   &= \left\{\begin{aligned}2(-1)^{m+n}\frac{mn}{m^2-n^2},\, m\neq n\\
	 \frac{1}{2},\,  m= n\end{aligned}\right.\label{Coef2_explicit}
	\end{align}

\section{Infinitesimal topography solution of the CMS}\label{app:WTA}
For a ridge topography, we have that $r\rightarrow 0$  as $x\rightarrow\pm\infty$ and $h_0$ is chosen as the far-field depth. For a shelf, $h$ is assumed as a depth transition from $h_0-\epsilon\,r_-$ to $h_0-\epsilon\,r_+$, with $r\rightarrow r_{\pm}$ as $x\rightarrow\pm\infty$ and $r_++r_-=0$ and $h_0$ represents the mean depth. Using the Taylor expansion
\begin{align*}
    \frac{1}{h} = \frac{1}{h_0-\epsilon r} = \frac{1}{h_0} + \frac{1}{h_0^2}\epsilon r + \dots,
\end{align*}
 we note that the various $h$-dependent terms in \eqref{eq:CMS} scale as follows
\begin{align*}
    \frac{h_x}{h} &= -\epsilon r_x\left(\frac{1}{h_0} + \frac{1}{h_0^2}\epsilon r+\dots\right) = O(\epsilon), &
    \frac{h_x^2}{h^2} &= O(\epsilon^2)&\\
    \frac{h_{xx}}{h} &= -\epsilon r_{xx}\left(\frac{1}{h_0} + \frac{1}{h_0^2}\epsilon r+\dots\right) = O(\epsilon), &
    h\left(\frac{1}{h}\right)_{xx} & = \epsilon \frac{r_{xx}}{h_0}+ O(\epsilon^2)&
\end{align*}
Substituting the above expressions together with $\phi_m =\sum_{i=0}^K\epsilon^i\phi_m^{(i)}$ in \eqref{eq:CMS}, we obtain $\phi_m^{(0)}=0$ and  \eqref{eq:CMS_WTA} for $\phi_m^{(1)}$. The  solution of \eqref{eq:CMS_WTA} satisfying the radiation conditions \eqref{eq:rad_cond_left} with $k_n^{\pm}=\ell_n$ is 
\begin{align}\label{eq:phin_WTA}
\phi^{(1)}_{n}= \frac{g_n}{\rev{h_0}}\left[ A_n(x;r)\ee{\ci \ell_n x} - B_n(x;r)\ee{-\ci \ell_n x}\right],\quad n\geq 1,
\end{align}
with
\begin{align}
  A_n(x;r)=\int_{-\infty}^{x}\ee{-\ci\ell_n s}r_s(s) \dd{s},\,\, B_n(x;r)=\int_{x}^{+\infty}\ee{\ci\ell_n s}r_s(s) \dd{s};
\end{align}
see, e.g., \citet[Chapter 7]{GZ08}). The response field at first order is reconstructed by \eqref{eq:RL_representation} with $\phi_n$ given by \eqref{eq:phin_WTA} and $Z_n$ by \eqref{eq:RL_vertical_function} with $h=h_0$. The conversion rate is obtained using \eqref{eq:phin_WTA} and \eqref{eq:Conv}, 
\begin{align}\label{eq:convrateWTA}
    C^{\text{WTA}}_{\pm} = \frac{\omega^2-N^2}{2\omega}\frac{h_{0}}{2}\sum\limits_{n=1}^{\infty}\Im\left\{\phi^{(1)}_n\rev{\overline{\phi^{(1)}_{n,x}}}\right\},\quad x\rightarrow\pm\infty.
\end{align}
In order to compute $C^{\text{WTA}}_-$, we first note that as $x\rightarrow -\infty$, we have $A_n\rightarrow 0$ and 
\begin{align*}
\phi^{(1)}_{n} &
         = -\frac{g_n}{\rev{h_0}}
         B_n(-\infty;r)\, \ee{-\ci \ell_n x}\\
         &= -\frac{g_n}{\rev{h_0}}\left(\int_{-\infty}^{\infty}\ee{\ci\ell_n s}r_s(s) \dd{s}\right)\, \ee{-\ci \ell_n x}\\
         & = -\frac{g_n}{\rev{h_0}}\left(\left[\ee{\ci\ell_n s}r(s)\right]_{-\infty}^{+\infty} -\ci\ell_n\int_{-\infty}^{\infty}\ee{\ci\ell_n s}r(s) \dd{s}\right)\ee{-\ci \ell_n x}\\
         & = -\frac{g_n}{\rev{h_0}}\left(\left[\ee{\ci\ell_n s}r(s)\right]_{-\infty}^{+\infty} -\ci\ell_n\overline{\hat{r}}(\ell_n)\right)\ee{-\ci \ell_n x},
\end{align*}
where $\hat{r}(\xi) = \int_{-\infty}^{+\infty}\exp(-\ci x\xi)r(s) ds$ is the Fourier transform of $r$ and $\hat{r}(-\xi) = \overline{\hat{r}}(\xi)$, since $r\in\mathbb{R}$. Similarly, we find
\begin{align*}
\phi^{(1)}_{n,x} &= \frac{g_n}{\rev{h_0}} \ci\ell_n B_n(-\infty;r)\, \ee{-\ci \ell_n x} \\
&= \frac{g_n}{\rev{h_0}}\left(\ci\ell_n\left[\ee{\ci\ell_n s}r(s)\right]_{-\infty}^{+\infty} +\ell_n^2\overline{\hat{r}}(\ell_n)\right)\ee{-\ci \ell_n x},\quad x\rightarrow-\infty.
\end{align*}
Combining the above results, we compute 
\begin{align*}
\phi^{(1)}_n\overline{\phi^{(1)}_{n,x}}   =-\frac{g_n^2}{\rev{h_0^2}}\Pi_n,\quad x\rightarrow-\infty,
\end{align*}
where
\begin{align*}
    \Pi_n   &=\left(\left[\ee{\ci\ell_n s}r(s)\right]^{+\infty}_{-\infty}-\ci\rev{\ell_n}\overline{\hat{r}}(\ell_n)\right)\left(-\ci\ell_n\left[\ee{-\ci\ell_n s}r(s)\right]^{+\infty}_{-\infty}+\ell_n^2\hat{r}(\ell_n)\right)\\
    &=-\ci\ell_n\left[r^2\right]^{+\infty}_{-\infty} + \left[\ee{\ci\ell_n s}r\right]^{+\infty}_{-\infty} \ell_n^2\hat{r}(\ell_n) + \ell_n^2\overline{\hat{r}}(\ell_n)\left[\ee{-\ci\ell_n s}r\right]^{+\infty}_{-\infty}-\ci\ell_n^3\overline{\hat{r}}(\ell_n)\hat{r}(\ell_n)\\
 &=-\ci\ell_n\left[r^2\right]^{+\infty}_{-\infty} + \left[\ee{\ci\ell_n s}r\right]^{+\infty}_{-\infty} \ell_n^2\hat{r}(\ell_n) + \overline{\ell_n^2\hat{r}(\ell_n)\left[\ee{\ci\ell_n s}r\right]^{+\infty}_{-\infty}}-\ci\ell_n^3\overline{\hat{r}}(\ell_n)\hat{r}(\ell_n)\\
 & = -\ci\ell_n\left[r^2\right]^{+\infty}_{-\infty} + 2\Re\left\{\left[\ee{\ci\ell_n s}r\right]^{+\infty}_{-\infty}\ell_n^2\hat{r}(\ell_n)\right\}-\ci\ell_n^3\overline{\hat{r}}(\ell_n)\hat{r}(\ell_n)
\end{align*}
The first term of the above right hand side vanishes for both the ridge and the shelf cases. It then follows from \eqref{eq:convrateWTA} that 
\begin{align*}
    C^{\text{WTA}}_{-} = -\frac{N^2-\omega^2}{2\omega}\frac{h_{0}}{2}\sum\limits_{n=1}^{\infty}\frac{g_n^2}{\rev{h_0^2}} \ell_n^3\overline{\hat{r}}(\ell_n)\hat{r}(\ell_n).
\end{align*}
Repeating the same procedure for $x\rightarrow+\infty$, we obtain $C_+^{\text{WTA}}=-C_-^{\text{WTA}}$. Recalling that $g_n =  Q (-1)^{n\rev{+1}}/(n\pi) $ and $Q=Uh_0$, we easily find that $\mathcal{C}^{\text{WTA}} = C_+^{\text{WTA}}-C_-^{\text{WTA}}$ is given by \eqref{eq:ConversionStL}.

\section{Reconstruction of the free-surface elevation \label{app:free_surface}}
It is possible to reconstruct the free-surface elevation $\eta$ induced by ITs within the rigid-lid approximation by using the baroclinic surface pressure at $z=0$, $p_s(x,t) = p^{\#}(x,0,t)$; $\eta = p_s/ g$. To find $p_s$, we evaluate $p_x^{\#}$ and $p_{xx}^{\#}$ on $z=0$, using \eqref{eq:Relations},
\begin{align}
\left[p^{\#}_{x}\right]_{z=0} = \left[f v^{\#} -u^{\#}_{t}\right]_{z=0}:=g(x,t),\,\,
\left[p^{\#}_{xx}\right]_{z=0} = \left[f v^{\#}_x -u^{\#}_{xt}\right]_{z=0}:=F(x,t),
\end{align}    
and we formulate and solve the following BVP on $[x_L,x_R]$,
\begin{align}\label{eq:fs_pert}
p^{\#}_{s,xx} = F(x,t),\quad
p^{\#}_{s,x}(x_L) = g(x_L,t),\quad
p^{\#}_{s,x}(x_R) = g(x_R,t).
\end{align}

\singlespacing
\bibliographystyle{agsm}
\bibliography{refs.bib}

\end{document}